\documentclass[useAMS,usenatbib,usegraphicx]{mn2e}
\usepackage{aas_macros}
\title[Inflow History and Chemical Evolution of M33]
{The Stellar Populations of M33's Outer Regions IV: 
Inflow History and Chemical Evolution}

\author[Barker \& Sarajedini]
{Michael K. Barker$^1$\thanks{Present address: 
Institute for Astronomy, University of Edinburgh, Blackford Hill, 
Edinburgh UK EH9 3HJ; mkb@roe.ac.uk} 
and A. Sarajedini$^1$\thanks{ata@astro.ufl.edu}\\
$^1$Department of Astronomy, University of Florida, Gainesville, FL 32611}

\begin{document}

\date{Accepted 2008 August 11.  Received 2008 July 13; in
  original form 2008 May 1}

\pagerange{\pageref{firstpage}--\pageref{lastpage}} \pubyear{2008}

\defcitealias{Eggen62}{ELS62}
\defcitealias{Searle78}{SZ78}
\defcitealias{Pagel95}{PT95}
\defcitealias{Pagel98}{PT98}
\defcitealias{Barker07a}{Paper II}
\defcitealias{Barker07b}{Paper III}

\maketitle

\label{firstpage}

\begin{abstract} 
We have modelled 
the observed color-magnitude diagram (CMD) 
at one location in M33's outskirts
under the framework of a simple chemical evolution
scenario which adopts instantaneous and delayed recycling
for the nucleosynthetic products of 
Type II and Ia supernovae.
In this scenario, interstellar gas forms stars at a rate
modulated by the Kennicutt-Schmidt relation 
and gas outflow occurs at a rate proportional 
to the star formation rate (SFR).
With this approach, we put broad constraints on the role of gas 
flows during this region's evolution and 
compare its [$\alpha$/Fe] vs.\ [Fe/H] relation
with that of other Local Group systems.
We find that models with gas inflow are significantly
better than the closed box model at reproducing
the observed distribution of stars in the CMD.
The best models have
a majority of gas inflow taking place in the last 7 Gyr, 
and relatively little in the last 3 Gyr.
These models predict most stars in
this region to have [$\alpha$/Fe] ratios lower than
the bulk of the Milky Way's halo.  
The predictions for the present-day SFR, gas mass, and 
oxygen abundance compare favorably to independent
empirical estimates.
Our results paint a picture in which M33's outer disc
formed from the protracted inflow of gas over several Gyr 
with at least half of the total inflow
occurring since $z \sim 1$.
\end{abstract}


\section{Introduction}
\label{sec:intro}

The distribution of stars in a color-magnitude diagram (CMD)
is a sensitive function of the past and present star formation
rate and chemical composition.
This fact allows the extraction of the star formation history (SFH) 
and chemical enrichment history (CEH) of a galaxy by fitting
its observed CMD with a model CMD whose SFH and CEH are known.
The results of this synthetic CMD fitting method can 
ultimately yield insights into the processes 
shaping galaxy evolution, such as gas flows, energetic feedback, 
satellite accretion, tidal interactions, and dynamical mixing.  
Nevertheless, this method is only one of many employed in
the domains of near-field cosmology and galactic paleontology 
\citep{Freeman02}.  

Another, equally important method, is 
chemical modelling, whose primary goal is to reproduce the 
elemental abundance distribution in a galaxy's stars.  
This method involves tracking the effects of gas flows, star formation, and
stellar nucleosynthesis on the abundances and abundance ratios
of certain elements.
For example, the abundance of the $\alpha$-elements
(O, Ne, Mg, Si, S, Ca, Ti) relative to iron, 
commonly plotted as [$\alpha$/Fe] vs.\ [Fe/H],
is an extremely useful tool to discriminate between 
different models for a galaxy's evolution because
the $\alpha$-elements and iron have different 
production sites and time-scales.  The $\alpha$-elements
are produced mainly in the hydrostatic burning phases 
and explosive deaths (supernovae Type II, or SNe II) 
of massive stars ($M \ga 8\ M_{\sun}$), 
which have short lifetimes ($\la 10^7$ yr).
The majority of iron, 
on the other hand, is created in supernovae Type Ia 
(SNe Ia), which are the thermonuclear explosions of carbon-oxygen 
white dwarfs in binary systems and, therefore, typically 
occur on a much longer time-scale of
$\sim 1$ Gyr \citep{Greggio05}.

The methods of CMD fitting and chemical modelling are complementary,
but they have evolved mostly independently.
Some examples in the recent literature which 
combine the two methods generally 
use the results of CMD fitting as external inputs to 
chemical evolution models.
For example, 
\citet{Aparicio97} derived the first SFH 
of the Local Group (LG) dwarf
galaxy, LGS 3, from CMD fitting and 
used the chemical evolution equations to 
see how much gas inflow and outflow was required to 
match their results.
\citet{Carigi02} computed the chemical evolution of
four dwarf spheroidal (dSph) satellites of the Milky Way (MW).   As external
constraints, these authors used the SFHs of these galaxies derived by 
\citet{Hernandez00} from a CMD fitting method.  
\citet{Dolphin02} calculated the SFHs of 6 dSph MW satellites 
using synthetic CMD fitting, and then 
\citet{Lanfranchi03,Lanfranchi04} adopted his SFHs 
as inputs to their chemical evolution models for the same galaxies.
Similarly, \citet{Fenner06} modelled the chemical evolution
of the Sculptor dSph using the SFH derived by 
\citet{Dolphin05} from a CMD analysis.
Lastly, after adopting the SFH derived by \citet{Wyder01,Wyder03},
\citet{Carigi06} modelled the evolution of 
NGC 6822 in a cosmological framework.

Conversely, the results of chemical modelling can be
used as external inputs to CMD fitting.  For example, 
\citet[][hereafter PT98]{Pagel98} modelled
the chemical evolution of the Large Magellanic Cloud (LMC) 
and Small Magellanic Cloud (SMC), 
adopting gas inflow and non-selective galactic winds 
(i.e., the wind efficiency was the same for all chemical elements).
These authors tuned the model parameters to match the observed
elemental abundances of clusters and supergiant stars in these systems.
Some subsequent studies adopted the LMC's 
age-metallicity relation (AMR) 
derived by \citetalias{Pagel98}
to extract its SFH from the CMD and luminosity function 
\citep[e.g.,][]{Holtzman99,SmeckerHane02}.

All the aforementioned studies combined CMD fitting and chemical modelling
in a two-step process, in which the first step was done
independently of the second.  
However, the two steps are inextricably linked because
an increase in the star formation rate (SFR) speeds
up the chemical enrichment.
Gas flows into or out of the system can change this 
coupling making them important ingredients to include in any model.
Therefore, one drawback of using the CMD-derived SFHs
as independent constraints on the chemical evolution models 
is that the SFHs are not necessarily physically 
self-consistent under the action of processes like
nucleosynthesis, galactic winds, and gas accretion,
processes which are fundamental to the chemical 
models themselves.  Similarly, using the results of
chemical evolution models as independent constraints on 
CMD fitting is not necessarily self-consistent because
the models have a particular form for the SFH, which is
the very thing the CMD fitting is supposed to derive.

The recent works of \citet{Ikuta02} and \citet{Yuk07} represent a significant
improvement over the studies mentioned above because they incorporate
CMD fitting and chemical modelling simultaneously.
\citet{Ikuta02} computed
a few closed box evolution models for several dSph satellites of the MW.
They performed a qualitative comparison of their 
model CMD and [Mg/Fe] vs.\ [Fe/H] 
relation with what was observed 
and found a reasonable agreement, but they had 
to invoke gas stripping
via ram pressure or tidal shocks to reconcile the
present day gas fraction of their closed box models ($\sim 97\%$)
with the observed values ($\sim 0\%$).
\citet{Yuk07} improve upon the work of \citet{Ikuta02}
by quantitatively fitting a closed box chemical 
model to the CMD of IC 1613, 
a relatively isolated LG dwarf irregular galaxy.
Their model SFH and AMR are in good agreement with
previous independent determinations based on the canonical
CMD fitting method, lending support to both the old and new methods.
Moreover, their predicted present-day oxygen abundance is
consistent with the observed value.

There are many lines of evidence that suggest 
not all galaxies evolve as closed boxes.
As originally hypothesized by \citet{Larson74} and
exemplified by the \citet{Ikuta02} results, gas outflows could
explain the lack of gas in dSphs despite their
low metallicities (see also \citet{Lanfranchi04}). 
Other evidence for gas outflows includes the 
abundances of metals in the IGM \citep{Edge91,Mushotzky97}, 
the mass-metallicity relation
of galaxies at low and high redshift 
\citep[e.g.,][]{Garnett02,Tremonti04,Pilyugin04,Erb06}, 
extended extra-planar HI gas in spirals 
\citep{Fraternali04,Oosterloo07}, 
high velocity clouds (HVCs) around the MW 
with near solar metallicity 
\citep{VanWoerden04,Wakker01,Richter01}, 
extra-planar optical and X-ray emission around starburst galaxies 
\citep{Heckman90,Lehnert96,Martin02,Strickland04}, and 
velocity shifts of high-ion absorption lines in 
damped Lyman-$\alpha$ systems 
\citep{Fox07} and 
Lyman break galaxies at $z \sim 3 - 4$ 
\citep[e.g.,][]{Pettini01,Adelberger03}.  For a more
detailed review, we refer the interested 
reader to \citet{Veilleux05}.

Evidence for gas inflows includes the so-called G-dwarf problem, 
which is the fact that the 
metallicity distribution function
of low mass, long lived stars observed in the Solar vicinity (SV) and
in many other galaxies is too narrow and contains too few
metal poor stars compared to the closed box model 
\citep[e.g.,][]{Tinsley75,RochaPinto96,Seth05,Jorgensen00,Wyse95,
Harris02,Koch06,Sarajedini05,Mouhcine05,Worthey05}.  
Additionally, gas inflow over several Gyr is required to 
explain many characteristics of the SV, including 
stellar chemical compositions 
and the present-day gas mass fraction and SFR 
\citep[e.g.,][]{Chiappini01,Portinari98}.  Other evidence
for inflows includes the kinematics of extra-planar HI gas in some spirals
\citep{Fraternali08}, the low metallicities and large
distances of some MW HVCs
\citep{Wakker99,Richter01}, 
high velocity OVI absorption along various sight-lines 
through the MW's halo \citep{Sembach03}, 
and the prevalence of warps in HI discs \citep{Bosma91}.
We refer the reader to \citet{Sancisi08} for a review
of observational evidence for gas inflow.

There are theoretical expectations for gas inflows, as well.
Cosmological simulations of galaxy
formation in the $\Lambda$CDM framework predict disc galaxies
to form from the accretion of gas after the last major
merger \citep{Abadi03,SommerLarsen03,Governato04,Governato07}.
For galaxies with virialized dark halo masses 
$\la 10^{12} M_{\sun}$, the accreted gas is expected
to be mostly cold ($T \sim 10^{4-5} K$).
On the other hand, 
upon entering the haloes of more massive galaxies, 
most of the accreted gas 
experiences shock-heating 
to the virial temperature ($T \ga 10^6 K$), creating 
surrounding reservoirs of hot gas \citep{Keres05,Dekel06}.
Thermal instabilities lead to the condensation of cooler
clouds, which then rain down on the discs 
\citep{Maller04,SommerLarsen06,Kaufmann06}.
In some of the most recent N-body simulations, 
these clouds have properties similar
to the HVCs mentioned above \citep[][]{Peek08}.
Furthermore, this idea is supported by the recent discovery of a hot 
($T \sim 10^6 K$) gaseous halo around 
the quiescent massive spiral, NGC 5746 
\citep{Pedersen06}.  This gas is too hot
to be heated by SNe in the disc and the disc SNe rate
is too low to have created the reservoir through the outflow of disc gas.
Finally, theoretical simulations suggest the need for extended accretion
of dilute gas to keep discs from being destroyed after a 
succession of minor mergers with mass ratios of 
4:1 and even up to 10:1 \citep{Bournaud07}.

Despite all this evidence, 
the precise nature and importance 
of gas flows in the evolution of galaxies, 
particularly spirals, is still uncertain.
To help improve the situation, in the present paper, 
we develop a chemophotometric
CMD fitting method as an extension to the canonical 
method used in many other studies, but 
with the goal of examining the
role of gas flows in M33's evolution.  Following 
\citet{Ikuta02}, we solve the chemical evolution equations
to obtain a self-consistent SFH and AMR.  
We improve upon their work by allowing for gas inflow and outflow
and by efficiently 
searching the full volume of parameter space to make
a detailed and quantitative fit to the observed CMD.

The photometric data we use in this paper and the reduction
procedure were presented in 
\citet[][hereafter \citetalias{Barker07a}]{Barker07a}.
In summary, three co-linear fields located in projection 
$\sim 20 - 30\arcmin$ southeast of M33's nucleus
were observed with the Advanced Camera for Surveys
on board the {\it Hubble Space Telescope}.  
In \citet[][hereafter \citetalias{Barker07b}]{Barker07b}, we
computed the SFHs for these fields using the canonical 
synthetic CMD fitting method with age and metallicity
as free parameters.
Because the outer two fields may have a non-negligible contribution
from M33's halo or thick-disc (see 
\citetalias{Barker07b} for a discussion), 
we restrict ourselves to
analyzing the innermost field in the current study.
This field has a projected galactic area of $\sim 0.7\ \rm kpc^2$ 
and it lies at a deprojected radius of $R_{dp} \sim 6$ disk scale lengths
\citep{Ferguson07} or 
$\sim 9$ kpc, assuming a distance
of 867 kpc \citep{Tiede04}, inclination of $56\degr$, 
and position angle of $23\degr$ \citep{Corbelli97}.
At this radius in M33, the azimuthally averaged HI 
and stellar surface densities are, respectively,
$\sim 3\ \rm M_{\sun}\ pc^{-2}$ and $\sim 0.3\ \rm M_{\sun}\ pc^{-2}$
\citep{Corbelli97,Corbelli03}.

In the next section, we outline the chemical evolution
equations, which form the backbone of our models.  
We describe in \S \ref{sec:method} how we link these equations with the
synthetic CMD fitting method 
to build a self-consistent model CMD.
In \S \ref{sec:chemresults}, we present the results
of fitting closed box and inflow/outflow models
to the data.  
We explore the effects of varying the model
parameters in \S \ref{sec:var}.
In \S \ref{sec:obs}, we test the model predictions
with independent observations 
and, in \S \ref{sec:chemdisc}, we compare 
the [$\alpha$/Fe] vs.\ [Fe/H] relation
in M33 with other LG systems.  
Finally, we summarize our results in \S \ref{sec:chemconc}.

\section{Chemical Evolution Equations}
\label{sec:chem}

To model the SFH of M33 in a self-consistent way we 
use the chemical evolution equations, which follow
the variation of the gas and stellar masses and the
abundances of various elements.
We use the instantaneous 
recycling approximation (IRA) in which stellar
lifetimes are treated as negligible compared to the age
of the Universe for masses greater than the present-day
main sequence turnoff mass ($M \ga 0.8\ M_{\sun}$).  
This is a good approximation for following 
the gas and stellar masses and the abundances of 
elements like oxygen, whose primary 
producers are massive stars.  
However, the IRA breaks down for elements with 
significant contributions from SNe Ia
because they are released
to the ISM on longer time-scales.
To follow the evolution of SNe Ia products we adopt the
delayed production approximation (DPA) of 
\citet[][hereafter PT95]{Pagel95}.  
In the DPA, 
there is a delay time, $\tau_d$, between the birth of a
stellar generation and the resulting SNe Ia explosions.

Our models track the elements O, Mg, Si, Ca, Ti, and Fe.  
To minimize the effect of uncertainties in the yield of 
any one particular $\alpha$-element, we focus on their 
sum, which we refer to as $\alpha$.
Unless otherwise noted, 
[$\alpha$/Fe] = [(O+Mg+Si+Ca+Ti)/(5 Fe)].
We do not follow carbon and nitrogen because they 
have a significant contribution from long lived low and
intermediate mass stars, for which the IRA and DPA break down.
In the future we plan to drop the IRA and DPA so that
we can follow these two elements.

We define the total mass surface density as 
$\Sigma(t) = \Sigma_s(t) + \Sigma_g(t)$
where $\Sigma_s(t)$ and
$\Sigma_g(t)$ are the stellar and gaseous surface densities,
respectively, and $t$ is the elapsed time of the model.  
We also define $\Sigma_i(t)$ as the gas mass 
surface density in the form of element $i$.
Combining the formalism of \citet{Tinsley80} 
and \citetalias{Pagel95}, the equations of
chemical evolution under the IRA and DPA can be expressed as
\begin{equation}
\frac{d\Sigma}{dt} = f_I(t) - f_O(t)
\label{equ:total}
\end{equation}
\begin{equation}
\frac{d\Sigma_s}{dt} = (1-R)\psi(t)
\label{equ:stars}
\end{equation}
\begin{equation}
\frac{d\Sigma_g}{dt} = -(1-R)\psi(t) + f_I(t) - f_O(t)
\label{equ:gas}
\end{equation}
\begin{eqnarray}
\frac{d\Sigma_i}{dt} &=& -(1-R)X_i(t)\psi(t) + y_i(1-R)\psi(t)\nonumber \\
& + & y_{d,i}(1-R)\psi(t-\tau_d) + f_{I,i}(t) - f_{O,i}(t)
\label{equ:elements}
\end{eqnarray}
where $\psi$ is the SFR, $f_{I}$ is the inflow rate, 
$f_{O}$ is the outflow rate, and $X_i = \Sigma_i / \Sigma_g$ is
the fraction of gas in the form of element $i$.  
The net stellar yield, $y_i$, is the mass of newly synthesized
element $i$ instantaneously returned to the ISM 
by a stellar generation per unit mass locked up into 
stellar remnants.  The delayed yield, $y_{d,i}$, 
represents the delayed chemical yield of SNe Ia.  
The return fraction, $R$, is the
fraction of mass returned to the ISM by a stellar generation.  
We calculate $R$ from the 
yields of \citet{vandenHoek97} 
(for the case of a variable mass loss efficiency) and \citet{Portinari98}.  
We find that, averaged over all metallicities, 
$R = 0.235$ for a \citet{Kroupa93} initial mass function (IMF).
We note in passing that the precise value of $R$ is
inconsequential because the factor $(1-R)$ can be absorbed into the
star formation efficiency parameter, $\epsilon$, described below.
However, we explicitly include $R$ to be more consistent
with previous studies.

\begin{table}
\begin{minipage}{\textwidth}
\caption{Adopted chemical yields and solar abundances}
\label{tab:yields}
\begin{tabular}{@{}lccc}
\hline
Element & $y_i$   &$y_{i,d}$& $X_{i,\sun}$\footnote{\citet{Anders89}} \\
\hline
O       & $6.68 \times 10^{-3}$ & $0.00$ & $9.546 \times 10^{-3}$ \\
Mg      & $4.53 \times 10^{-4}$ & $0.00$ & $5.144 \times 10^{-4}$ \\
Si      & $4.56 \times 10^{-4}$ & $7.82 \times 10^{-5}$ & $6.518 \times 10^{-4}$ \\
Ca      & $3.36 \times 10^{-5}$ & $1.08 \times 10^{-5}$ & $6.007 \times 10^{-5}$ \\
Ti      & $1.49 \times 10^{-6}$ & $2.56 \times 10^{-7}$ & $2.130 \times 10^{-6}$ \\
Fe      & $3.23 \times 10^{-4}$ & $4.85 \times 10^{-4}$ & $1.155 \times 10^{-3}$ \\
\hline
\end{tabular}
\end{minipage}
\end{table}

Equs.\ \ref{equ:total} -- \ref{equ:elements} 
represent conservation of total, gaseous, stellar, and elemental mass.
The first term in Equ.\ \ref{equ:elements} represents the mass 
of element $i$ that is originally present in the gas
and lost to star formation, plus what is instantaneously returned
by the winds and explosive deaths of massive stars.  The second
term in Equ.\ \ref{equ:elements} represents the 
instantaneous restitution
rate of newly synthesized element $i$ while the 
third term is the delayed restitution of newly 
synthesized element $i$ (from SNe Ia), which is 
zero for $t < \tau_d$.  
We adopt the semi-empirical yields (shown in Table \ref{tab:yields}) and
time delay ($\tau_d = 1.3$ Gyr) used 
by \citetalias{Pagel95}, which they tuned to reproduce the observed
abundances and gas fraction in the SV and then applied 
to the LMC and SMC \citepalias{Pagel98}.
The fourth and fifth terms of Equ.\ \ref{equ:elements}
represent, respectively, the inflow and outflow rate of
element $i$.  The chemical
composition of the inflowing gas is assumed primordial.
The initial heavy element abundance and stellar mass 
are zero and, except for closed box models, 
the initial gas mass is zero, also.

\begin{table*}
\begin{minipage}{126mm}
\caption{Fit qualities, distances, and extinctions of the Padova solutions}
\label{tab:results1padova}
\begin{tabular}{@{}lrrrrrrrrrr}
\hline
Model & $Q$ & $\chi^2_{\nu}$ & $\nu$ & $\overline{(m-M)_0}$ & $\sigma$
& $\overline{A_V}$ & $\sigma$ \\
\hline
Closed box  &   29.13  &    2.42  & 1743 &   24.50  &    0.05  &    0.25  &    0.03\\
Ex. inflow  &   19.28  &    2.03  & 1741 &   24.56  &    0.07  &    0.21  &    0.05\\
Sand. inflow  &   15.37  &    1.99  & 1741 &   24.60  &    0.05  &    0.18  &    0.04\\
Double Ex. inflow  & 9.16  &    1.81  & 1739 &   24.60  &    0.05  &    0.18  &    0.04\\
Truncated inflow  &  8.68  &    1.78  & 1739 &   24.60  &    0.05  &    0.20  &    0.06\\
Free inflow  &   8.49  &    1.77  & 1739 &   24.60  &    0.05  &    0.20  &    0.06\\
\hline
\end{tabular}
\end{minipage}
\end{table*}

\begin{table*}
\begin{minipage}{126mm}
\caption{Star formation and outflow efficiencies of the Padova solutions}
\label{tab:results2padova}
\begin{tabular}{@{}lrrrrrrrrrr}
\hline
Model & $\overline{\rm log(\epsilon)}$ & $\sigma_{\rm hi}$ &
$\sigma_{\rm lo}$ & $\overline{w}$ & $\sigma_{\rm hi}$ & $\sigma_{\rm lo}$ \\
\hline
Closed box  &   -2.29  &    0.11  &    0.05  & ...  & ...  & ...\\
Ex. inflow  &   -1.43  &    0.72  &    0.44  &    0.66  &    1.11  &    0.52\\
Sand. inflow  &   -0.69  &    0.76  &    0.64  &    0.99  &    0.67  &    0.52\\
Double Ex. inflow  &   0.77  &    0.48  &    1.28  &    1.44  &    0.47  &    0.80\\
Truncated inflow  &   -0.47  &    0.55  &    0.58  &    0.98  &    0.55  &    0.56\\
Free inflow  &   -0.21  &    1.05  &    0.65  &    1.21  &    0.75  &    0.78\\
\hline
\end{tabular}
\end{minipage}
\end{table*}

Motivated by the successes of chemical
evolution models applied to the MW's satellites
\citep[e.g.,][]{Lanfranchi04},
we set the outflow rate proportional to the SFR 
by a constant factor, $w$, called
the outflow efficiency.  
The composition of the 
outflowing gas is the same as the ISM, which is
assumed to be homogeneous and instantaneously mixed.  
In chemical evolution models of LG dwarf
galaxies the outflow efficiency is $\sim 1 - 10$
\citep{Lanfranchi04,Carigi06}, 
so we restrict $w$ to be less than 10.
The idea behind this approach
is that SNe inject kinetic energy into the ISM,
possibly causing some gas to leave the system entirely.
In reality, the outflow efficiency could depend on 
factors like the depth of the gravitational potential, 
the ejecta velocity, and 
the ISM's physical properties.
A more detailed calculation
including such properties is beyond the scope of
the present study.  
We make no explicit distinction between radial or extra-planar
gas flows.

As is commonly done in chemical evolution models, 
we couple the SFR to the gas mass through the 
so-called Kennicutt-Schmidt (KS) relation, 
\begin{equation}
\psi(t) = \epsilon\ \Sigma_g^{\kappa}(t),
\end{equation}
where $\epsilon$ is the star formation efficiency.
In the spirit of the pioneering work of \citet{Schmidt59}, 
who found that star formation rate traced gas density in 
the MW, \citet{Kennicutt98} measured mean gas masses and SFRs
within the optical radii of $\sim 60$ normal
spirals and within the central regions
of $\sim 30$ starburst galaxies and found 
$\epsilon = 0.25 \pm 0.07$ and 
$\kappa = 1.40 \pm 0.15$~
\footnote{on a scale where
$M_{\sun}\ {\rm pc^{-2}}\ {\rm Gyr^{-1}}$ are the units of $\psi$ 
and $M_{\sun}\ {\rm pc^{-2}}$ 
are the units of $\Sigma_g$}.
Looking at the two subsamples individually, 
he found $\kappa = 2.47 \pm 0.39$ for the normal spirals
and $\kappa = 1.40 \pm 0.13$ for the starbursts.

\begin{table*}
\begin{minipage}{126mm}
\caption{Fit qualities, distances, and extinctions of the Teramo solutions}
\label{tab:results1teramo}
\begin{tabular}{@{}lrrrrrrrrrr}
\hline
Model & $Q$ & $\chi^2_{\nu}$ & $\nu$ & $\overline{(m-M)_0}$ & $\sigma$
& $\overline{A_V}$ & $\sigma$ \\
\hline
Closed box  &   33.26  &    2.41  & 1743 &   24.63  &    0.08  &    0.17  &    0.04\\
Ex. inflow  &   22.61  &    2.18  & 1741 &   24.60  &    0.05  &    0.10  &    0.03\\
Sand. inflow  &   15.08  &    2.02  & 1741 &   24.70  &    0.05  &    0.10  &    0.03\\
Double Ex. inflow  &    9.26  &    1.90  & 1739 &   24.70  &    0.05  &    0.10  &    0.03\\
Truncated inflow  &    9.59  &    1.90  & 1739 &   24.70  &    0.05  &    0.10  &    0.03\\
Free inflow  &    8.93  &    1.89  & 1739 &   24.70  &    0.05  &    0.10  &    0.03\\
\hline
\end{tabular}
\end{minipage}
\end{table*}

\begin{table*}
\begin{minipage}{126mm}
\caption{Star formation and outflow efficiencies of the Teramo solutions}
\label{tab:results2teramo}
\begin{tabular}{@{}lrrrrrrrrrr}
\hline
Model & $\overline{\rm log(\epsilon)}$ & $\sigma_{\rm hi}$ &
$\sigma_{\rm lo}$ & $\overline{w}$ & $\sigma_{\rm hi}$ & $\sigma_{\rm lo}$ \\
\hline
Closed box  &   -1.75  &    0.16  &    0.15  & ...  & ...  & ...\\
Ex. inflow  &   -0.19  &    0.70  &    0.28  &    0.15  &    0.72  &    0.15\\
Sand. inflow  &    0.33  &    0.64  &    0.84  &    0.82  &    0.18  &    0.57\\
Double Ex. inflow  &    0.99  &    0.00  &    0.77  &    0.78  &    0.14  &    0.30\\
Truncated inflow  &    0.30  &    0.70  &    0.31  &    0.40  &    0.52  &    0.25\\
Free inflow  &    1.00  &    0.00  &    0.57  &    0.77  &    0.16  &    0.20\\
\hline
\end{tabular}
\end{minipage}
\end{table*}

This global star formation relation can be compared to
a local one which refers to gas densities and SFRs
at individual points or in azimuthally
averaged bins within a galaxy.
Several studies of nearby systems 
have found that, when examined locally, 
$\kappa \approx 1.2 - 3.5$ 
\citep{Gottesman77,Wong02,Kennicutt07,Boissier03,Misiriotis06}.
The cause of the observed variation in $\kappa$ is unclear but possibilities
include non-axisymmetric profiles or uncertainties in 
the extinction correction, the CO-$\rm H_2$ conversion factor, 
flux calibration, and the conversion to SFR.
The variations may be due to intrinsic
properties of the galaxies, as there is some evidence that
molecule-rich galaxies, typically massive spirals and starbursts, 
have lower $\kappa$ than molecule-poor galaxies, typically
low surface brightness and dwarf systems.  Interestingly, M33
falls into this latter category with a total molecular 
fraction of $\sim 0.1$ \citep{Corbelli03}.

The dependence of $\kappa$ on molecular fraction is 
not too surprising since
star formation is observed to trace molecular gas quite well 
\citep[e.g.,][]{Murgia02,Heyer04,Matthews05,Leroy05,Gardan07}.
Indeed, the studies mentioned above also found that, 
when considering the molecular gas alone, $\kappa \approx 1.4$, 
with much less galaxy-to-galaxy variation than
when considering the total or atomic gas.
It is beyond the scope of the present study to track the
time evolution of the molecular gas fraction, 
but this is one possible avenue for future improvement
to the techniques presented here.

With these considerations in mind, 
an appropriate empirical KS relation to use would be one
derived for M33 involving the total gas density.  
\citet{Heyer04} measured $\kappa = 3.3 \pm 0.07$ 
and $\epsilon = 0.0035 \pm 0.066$ 
using the infrared luminosity to estimate M33's SFR profile
inside $R_{dp} \sim 30\arcmin$.  
Similarly, \citet{Boissier07} studied M33's KS relation
by measuring its UV surface brightness
and translating that to a SFR profile.
Although \citet{Boissier07} did not report specific values, 
we estimate by eye from their figures 
a similar value for $\kappa$ as \citet{Heyer04}, 
but a smaller value for $\epsilon$ by a factor of $\sim 3$.
Adopting the \citet{Heyer04} values for $\epsilon$ and $\kappa$
leads to very poor fits of the CMD because
the distributions of stellar age and metallicity 
are skewed too high and low, respectively (see \S \ref{sec:exp}).
Further tests show that our models are more sensitive
to reasonable changes in $\epsilon$ than in $\kappa$, so we
adopt $\kappa = 3.3$ for all models and allow $\epsilon$ to be a free
parameter (constant in time) restricted to the range $0.0001 - 10.0$.
We note that
adopting $\kappa = 1.4$ does not significantly change
our general conclusions because the gas mass surface density
$\sim 1\ M_{\sun}\ \rm pc^{-2}$ 
throughout much of the system's evolution.
Finally, we do not include a 
star formation threshold 
(see the discussion in \citetalias{Barker07a} 
and \citet{Boissier07} for reasons why).

Our chemical models can be summarized as follows.
Gas inflow deposits gas into the system and 
drives star formation via the KS relation.
Because the inflowing gas has primordial composition, 
the ISM metallicity is always below what it would be without inflow.  
About $25\%$ of the gas that goes into making each
stellar generation is instantaneously returned to the ISM by
stellar winds and SNe II, while
$\sim 75\%$ is locked up into stellar remnants like
white dwarfs, neutron stars, and black holes.
The star formation enriches the ISM with metals, 
but also drives an outflow of gas
from the system.  
All else being equal, the presence of a gas outflow
quenches the SFR, slows the chemical enrichment, 
and suppresses the 
metallicity below what it would be without outflow.
The ISM metallicity generally increases with time
due to stellar nucleosynthesis, but 
a short, rapid increase in the inflow rate (IFR) can 
decrease the ISM metallicity temporarily.
In the absence of all gas flows, an initial gas
reservoir must be present to start SF, which 
continuously declines thereafter.

\section{Method}
\label{sec:method}

We start with a set of model parameters (e.g., $\epsilon$, $\kappa$, $w$,
and $f_I$), which we use to solve 
Equs.\ \ref{equ:total} $-$ \ref{equ:elements}.
We integrate these equations using a 4th-order
Runge-Kutta method with a time-step of $2 \times 10^{6}$ yr.  
The numerical integration was checked against several 
analytic solutions and the typical fractional error was $\sim 10^{-6}$.
The resulting chemical evolution model specifies the total and
gas mass surface densities, SFR, and abundances of 
the various elements as functions of time.

The age-metallicity plane is divided 
into logarithmic bins of width 0.25 dex in age and 0.3 dex in 
metal abundance.  To cover this plane, 
we use the same set of synthetic CMDs as 
in \citetalias{Barker07b}, 
each of which represents the predicted photometric
distribution of stars in the corresponding age and metallicity bins.
Each of these CMDs was created with 
the IAC-star program \citep{Aparicio04}.
We calculate the total stellar mass formed in each 
age bin from the chemical model and the 
corresponding mass-weighted 
mean global metallicity 
from Equ.\ \ref{equ:mh} described below.
If the metallicity falls outside the CMD library, 
it is capped at the appropriate limit.
In each age bin, the total stellar mass 
is split up into the two
synthetic CMDs which bracket the mean metallicity.
More weight is given to the CMD whose central 
metallicity is closer to the mean metallicity.
The final model CMD is the superposition of all 
synthetic CMDs weighted by their amplitudes.
The data and model CMDs are divided into square 
bins 0.1 mag on a side and compared using the
maximum likelihood ratio, $\Upsilon$, for a Poisson distribution.
The model parameters are changed and the whole process
repeats until $\Upsilon$ is minimized.

The quality of the fit is formally measured by the parameter, 
$Q$, which gives the difference between $\Upsilon$ and its
expectation value in units of its standard deviation.
A $Q$ value of 0.0 indicates a perfect fit, while, for example,
a value of 2.0 indicates a $2\sigma$ departure from
a perfect fit \citep[][\citetalias{Barker07b}]{Dolphin02}.
Because many readers may be more familiar with $\chi^2$ than $Q$,
we also calculate the reduced $\chi^2$ of
the best fit.
As in \citetalias{Barker07b}, we simultaneously solve for
distance and extinction with a fixed
grid spanning the ranges $(m-M)_0 = 24.50 - 24.80$ and 
$A_V = 0.10 - 0.25$ in steps of 0.10 and 0.05 mag, respectively.
The global best fit (referred to simply as the best fit) 
is the weighted average of those
distance/extinction combinations whose solutions 
lie within $1\sigma$ of the best individual solution.

We use the program, 
StarFISH \citep{Harris01}, after incorporating 
the genetic algorithm, PIKAIA \citep{Charbonneau95}, to 
efficiently search the full volume of parameter space.  
Briefly, this algorithm randomly generates an initial 
population of solutions, which is evolved through 
successive generations under the action of natural selection, 
breeding, inheritance, and random mutation to find the global solution.
We ran PIKAIA for 200 generations in the
steady-state-replace-random reproduction mode
with creep mutation enabled and 50 individuals in each
generation.
Then, the downhill simplex routine of StarFISH was
started from the best-fitting solution found by PIKAIA.
This hybrid approach proved more efficient at locating the
global solution than either method individually.

Because we are no longer solving for the amplitudes
of the basis populations, we had to modify
the way StarFISH calculates the confidence intervals.
For each parameter, we take small steps 
in the positive and negative directions 
away from its optimum value.
After each step, we allow the downhill simplex to re-converge 
while holding that parameter fixed and allowing the
others to vary.
The process is repeated until the 1$\sigma$ 
limit of the fitting statistic was reached.

To estimate the systematic errors in the stellar
evolutionary tracks we created two sets of synthetic
CMDs using the \citet{Girardi00} and 
\citet{Pietrinferni04} tracks, which we respectively 
designate as Padova and Teramo (also referred to as BaSTI in
the literature).
The conversion from the theoretical to the observational
plane is accomplished with the \citet{Castelli03}
library of bolometric corrections.

At this point it is important to note that these
stellar tracks have scaled-solar abundances, but
the elemental abundance distribution in forming stars changes
with time.  This is particularly important for the RGB,
whose temperature is determined mostly by the 
abundances of elements with low first ionization
potentials (Fe, Mg, Si, S, and Ca), which are the
primary atmospheric electron donors.
Ideally, the stellar tracks would exist for 
all possible elemental abundance
distributions and we would pick the appropriate
tracks at each time step of the chemical model.
However, the computational effort required to generate
such a comprehensive set of tracks is too 
prohibitive.  Only recently have groups begun to
compute tracks for many different masses and
metallicities and different levels of $\alpha$-enhancement, 
assigning similar enhancement
to all the $\alpha$-elements, but on a very coarse grid
in [$\alpha$/Fe] 
\citep[e.g.,][]{VandenBerg06,Pietrinferni06,Dotter07a}.

\citet{Salaris93} found that $\alpha$-enhanced 
tracks and isochrones are well reproduced by scaled-solar
ones with the same global metallicity provided the enhancements
are similar for all the $\alpha$-elements.
Subsequent studies demonstrated that this result 
begins to break down for $Z \ga 0.003$ 
\citep[e.g.][]{Salaris98,Salasnich00,VandenBerg00,Kim02b}.  
We have checked its applicability using 
several of the most recent isochrone databases
\citep{Pietrinferni06,Dotter07a,VandenBerg06} and we find that 
over the range of ages, [Fe/H], and [$\alpha$/Fe] 
most appropriate for our data, a scaled-solar isochrone
in the $I\ {\rm vs.}\ (V-I)$ plane 
is within $\sim 0.1$ mag of an $\alpha$-enhanced isochrone
with the same global metallicity.
Therefore, we account for $\alpha$-element enhancements
by calculating the global metallicity following the 
formalism of \citet{Salaris93} and \citet{Piersanti07}.
\begin{equation}
\rm [M/H] = [Fe/H] + log(a_{\alpha}10^{[\alpha/Fe]} + b_{\alpha})
\label{equ:mh}
\end{equation}
In Equ.\ \ref{equ:mh}, 
${\rm a_{\alpha}} = \sum_i(X_i/Z)_{\sun} \approx 0.7$ 
for $i =$ O, Ne, Mg, Si, S, Ca, and Ti 
and $\rm b_{\alpha} = 1 - a_{\alpha}$.
Note that this relation implicitly assumes no 
enhancements in elements other than the $\alpha$-elements. 
This is not a significant
problem, however, since the $\alpha$-elements comprise 
most of the metals by mass.
In any case, our conclusions are insensitive 
to the precise value of the correction factor 
to [Fe/H] in Equ.\ \ref{equ:mh} because it
generally amounts only to $< 0.2$ dex.

In principle, the bolometric 
corrections, color transformations, 
and evolutionary lifetimes also depend 
on [$\alpha$/Fe], but in practice, our data
are not significantly affected by these dependencies.  
\citet{Cassisi04} showed that
the bolometric corrections and color transformations 
depend negligibly on [$\alpha$/Fe] in $V$ and redder bands.
\citet{Dotter07b} investigated the effect of
abundance variations on stellar evolutionary models
and found that when $\rm [\alpha/Fe] = 0.3$, the MS lifetime
is decreased by $\sim 5\%$.  
We have compared the scaled-solar and $\alpha$-enhanced 
Teramo tracks, which have $\rm [\alpha/Fe] \sim 0.4$ 
\citep{Pietrinferni06}, and we find that, in general, the 
evolutionary phase lifetimes differ by $\la 10\%$.
These effects are likely to be even smaller 
for our data since we derive 
$\rm -0.2 \la [\alpha/Fe] \la 0.2$
for almost all ages.

\section{Results}
\label{sec:chemresults}

\begin{figure*}
\includegraphics[width=\textwidth]{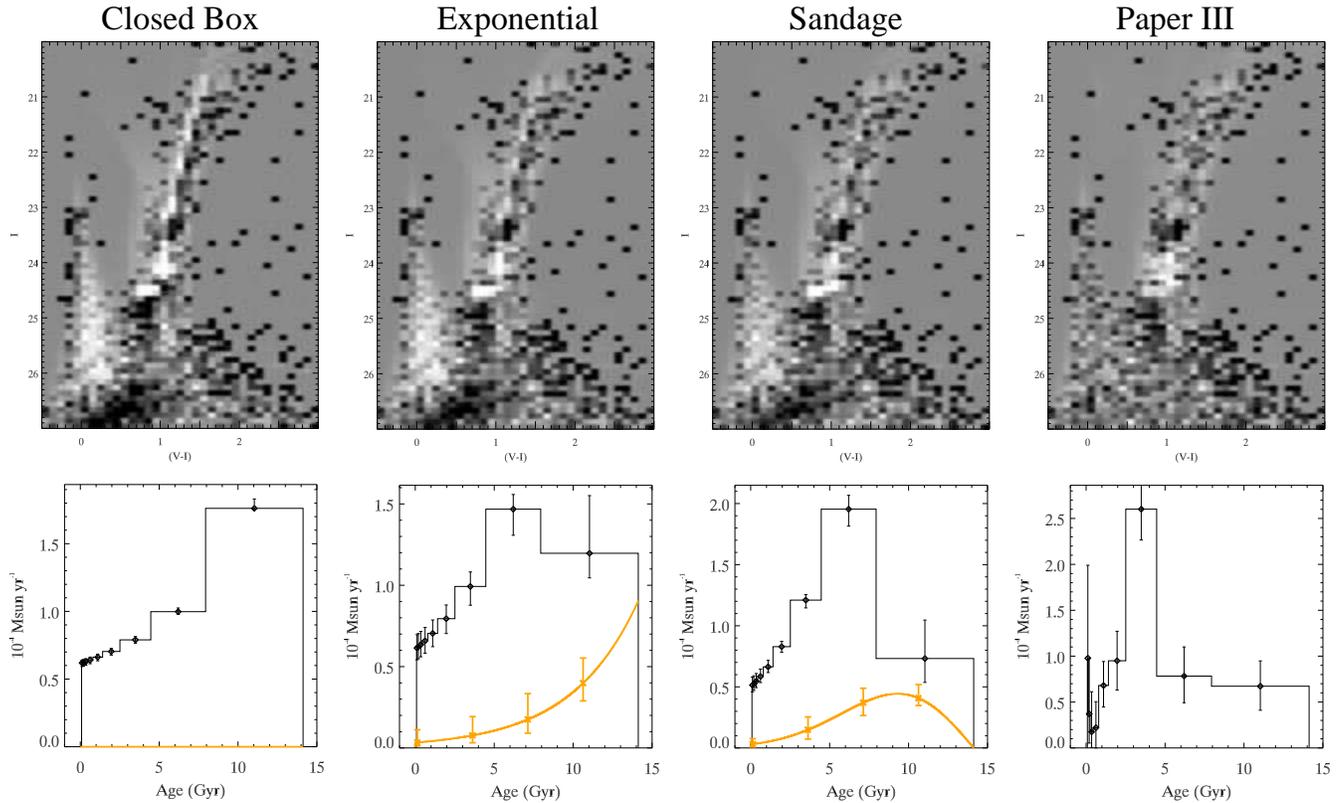}
\caption{Star formation history and inflow history 
results using the Padova tracks.  
The top panel in each column shows the residual CMD on a range 
of $\pm 3\sigma$ where positive (white) residuals indicate an excess
of model stars.
The bottom panel in each column shows the star formation history as the upper 
line and inflow history (scaled by a factor of 0.1 and integrated
over the total field area) as the lower line.
From left to right, the columns correspond to 
the closed box, exponential inflow, and Sandage inflow
models, respectively.  The last column shows the results
from \citetalias{Barker07b}, for which age and metallicity
were free parameters.}
\label{fig:A1_padova_ira_grid1}
\end{figure*}

We tested several different inflow/outflow scenarios
using both sets of stellar tracks.
In general, they give similar results, so 
we only show those of the Padova
tracks.  In the text, we describe any significant differences, since they
can help us gauge the systematic errors due to the tracks themselves.
The top row in Figs.\ \ref{fig:A1_padova_ira_grid1} -- 
\ref{fig:A1_padova_ira_grid2}
shows the residual CMD of a particular scenario on a scale where
white and black correspond, respectively, to an excess and deficit 
of model stars at the $3\sigma$ level.
The bottom row in these figures shows the best-fitting SFH
as the upper line and inflow history (IFH) divided by 10 as the lower line.
The IFH is integrated over the total field area.
The last column in these figures gives the result of
\citetalias{Barker07b}, in which age and metallicity were
free parameters.
Tables \ref{tab:results1padova} and \ref{tab:results1teramo} give the
the fit quality ($Q$), reduced $\chi^2$ ($\chi_{\nu}^2$), 
number of degrees of freedom ($\nu$), 
and mean distance modulus and extinction with their respective
$1\sigma$ uncertainties.
Tables \ref{tab:results2padova} and \ref{tab:results2teramo} give the
mean star formation and outflow efficiencies and
their upper and lower $1\sigma$ uncertainties.
The mean values reported in the tables are averages of the
acceptable solutions in the distance/extinction grid as explained in
\S 3.

\subsection{Closed Box Models}
\label{sec:closedbox}

We began by testing the canonical closed box model in which the
inflow and outflow rates were identically zero.
The system was initially composed entirely of gas. 
The total mass remained constant throughout its 
evolution while the gas mass and SFR decreased monotonically
with time.  The best-fitting closed box model is 
displayed in the first column of Fig.\ \ref{fig:A1_padova_ira_grid1}.

There are several significant discrepancies between
the model and data CMDs.
First, the model predicts too much
star formation at ages $\la 1$ Gyr which causes
an overabundance of model main sequence (MS) stars on the blue plume 
at $(V-I) \sim 0.0$.
Second, the model 
has too few stars in a region below the blue plume 
centred at $(V-I) \sim 0.5$ and $I \sim 26.5$.  
In the global solutions of \citetalias{Barker07b},
this region is dominated by
stars with ages $\sim 2 - 8$ Gyr, indicating the closed box
model has too little star formation at these ages.
Also, the RGB of the model is too blue, indicating
a mean metallicity that is too low.

The Teramo model exhibits similar discrepancies, but the 
RGB appears
slightly too wide, possibly indicating an excessively
large metallicity spread, and it has too many
stars overall.  
The Teramo model also has 
too many stars on the blue horizontal branch, which signals that
there is too much star formation at the oldest ages.
It cannot be due to the metallicity of the
oldest bin being too low because it is already close
to that of the \citetalias{Barker07b} solution, which does not
show such a discrepancy.

The fit qualities of the closed box solutions 
are $> 20\sigma$ worse than the solutions of 
\citetalias{Barker07b}, which had $Q$ values 
of 6.64 (Padova) and 6.02 (Teramo).  
This is not surprising since our closed box model
has only two free parameters, namely, the star formation
efficiency and the total mass.  However, as we will
see below, the addition of gas inflow and outflow significantly
improves the fits with only $2 - 4$ more free parameters, 
indicating this region in M33 probably did not
evolve as a closed box.
Because of the reduced number of free parameters, the
confidence intervals of the closed box 
solutions are significantly smaller 
than those of the \citetalias{Barker07b} solutions.
This demonstrates that simultaneous CMD and chemical evolution
modelling can be used to alleviate the age-metallicity
degeneracy inherent in broad-band stellar colors.

\subsection{Inflow and Outflow Models}
\label{sec:exp}

\begin{figure*}
\includegraphics[width=\textwidth]{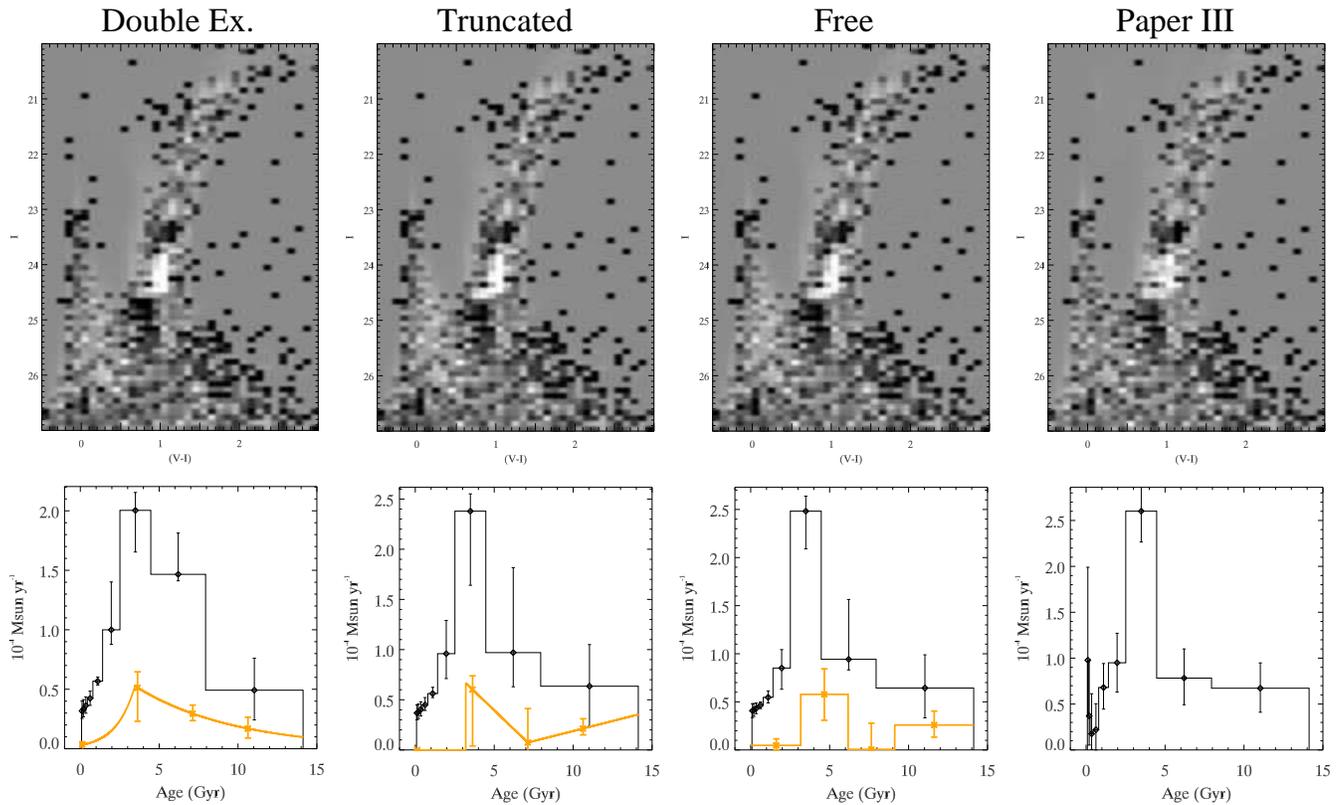}
\caption{Same as Fig.\ \ref{fig:A1_padova_ira_grid1}, but 
the first three columns show the double exponential, 
truncated, and free inflow models, respectively.}
\label{fig:A1_padova_ira_grid2}
\end{figure*}

As described in \S\ref{sec:intro}, there is 
evidence that galaxies do not evolve
as closed boxes.  An {\it exponential} inflow rate is one of the
simplest and most common forms used in the literature,
so it is instructive to see how well it can explain M33's SFH.  
Accordingly, we solved for 
the inflow time-scale and the initial inflow rate.
We also included gas outflow by allowing a nonzero 
outflow efficiency, $w$.
The second column in Fig.\ \ref{fig:A1_padova_ira_grid1} shows the
exponential inflow model for the Padova tracks.

This new model provides a better fit
to the data than the closed box model, but it still exhibits 
some large discrepancies with the data.  In fact, these
discrepancies are very similar to those of 
the closed box model, but the magnitude of the residuals 
has been lessened.  There is still too much
star formation at ages $\la 1$ Gyr and too little in the
range $2 - 8$ Gyr.
The fit quality is $\sim 15\sigma$ worse than the
\citetalias{Barker07b} solutions.

Some numerical simulations of structure formation
within the $\Lambda$CDM framework
predict that the average mass accretion rates of dark matter haloes 
are initially small, grow to some maximum, and 
decline thereafter \citep[e.g.,][]{vandenBosch02,Wechsler02}.
With this in mind, we also investigated another function
for the IFR which has a delayed maximum, 
$f_I(t) \propto t\ \exp(-t^2/2\tau^2)$,
where $\tau$ is the time between when the inflow starts and
when it peaks.  This function, which we refer 
to as {\it Sandage inflow}, was first used 
by \citet{Sandage86} to describe the variation in SFH with 
galaxy morphology and later explicitly presented
by \citet{MacArthur04}.
By shifting the bulk of the inflow toward younger ages
(i.e., earlier lookback times), 
it allows for more star formation at correspondingly
younger ages.
The result is plotted in the third column of
Fig.\ \ref{fig:A1_padova_ira_grid1}.  
The morphologies of the residuals are similar to the
exponential inflow model, but their magnitudes are smaller.
The fit quality is $\sim 9\sigma$ worse than the
\citetalias{Barker07b} solutions.

\begin{figure*}
\includegraphics[]{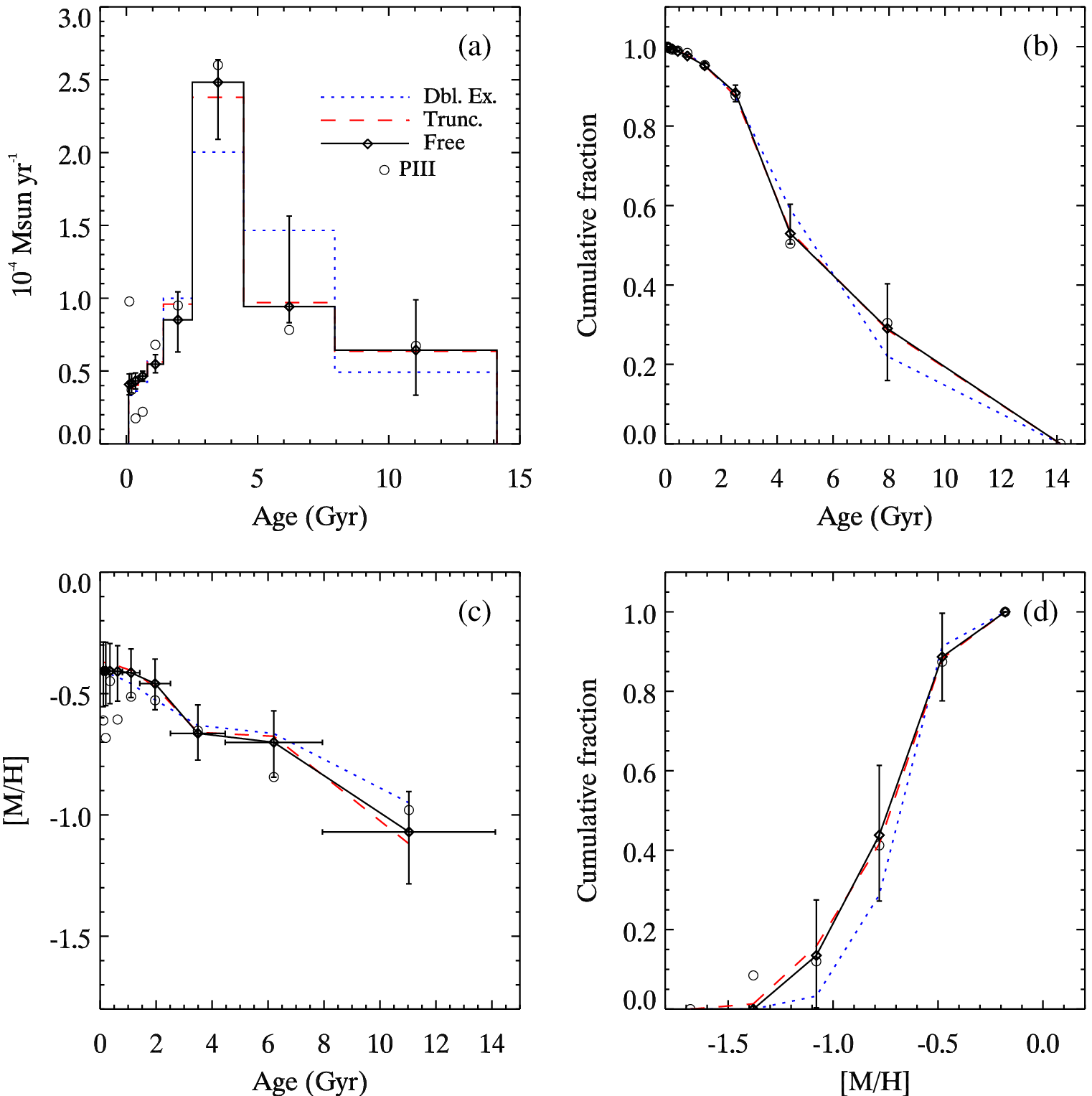}
\caption{(a) Star formation history, 
(b) age cumulative distribution function, 
(c) age-metallicity relation, and (d) metallicity distribution function 
of the acceptable inflow solutions using the Padova tracks.
Each panel shows the double exponential (dotted), 
truncated (dashed), and free inflow (solid) models
and the results from \citetalias{Barker07b} (open circles).
Only the free inflow model errors are shown for the sake of clarity.}
\label{fig:A1_padova_ira_compare_sfh}
\end{figure*}

The Teramo models show a qualitatively similar behavior --
the exponential and Sandage functions do a progressively
better job at reproducing the observed CMD, but still do not
get the overall age distribution correct.
The main drawback of these functions
is that the IFR today cannot easily be varied independently of the
IFR at intermediate ages ($\sim 2 - 8$ Gyr).  
These functions just do not provide enough freedom to
describe the true IFH which
may not be adequately characterized by just two parameters
or a smooth function.

These results led us to try three less restrictive inflow models.  
The first was a {\it double exponential} model described by four
parameters: a growing time-scale, a decaying time-scale, a
transition time between the growing and decaying modes,
and the IFR at the transition time.  The second was a
{\it truncated} model 
described by four parameters: the initial IFR,
the IFR at a model time of 7 Gyr, a truncation time when 
the inflow ends, and the IFR at the truncation time.
In the third model, which we called {\it free inflow}, 
we approximated the true IFH with a discrete 
function by dividing the 
entire age range into 4 bins each with a
possibly different but constant IFR.  
We experimented with different binning schemes 
and settled upon a compromise between
the desire to have a reasonable computing time and 
small error bars (requiring fewer bins)
and the competing desire to produce 
complicated SFHs (requiring more bins).
The free inflow model allows discontinuous
jumps in the IFR, but these should not be interpreted
too literally.  This is the same approach that we 
take in approximating the true SFH with 9 logarithmic 
bins, each with constant SFR.
Internal tests have shown that these inflow functions
can reasonably capture broad trends in the
true IFH and chemical evolution, especially at ages $\la 7$ Gyr.

The solutions using these three inflow models 
are displayed in 
Fig.\ \ref{fig:A1_padova_ira_grid2}.
The free inflow 
model provides the highest quality fits.
However, the double exponential and truncated inflow
models are $< 1\sigma$ worse and exhibit
qualitatively similar results.
The discrepancies between model and data 
exhibited previously with the exponential and Sandage inflow models 
are almost completely erased.  
Most of the discrepancies that do remain are similar
to those exhibited by the solutions 
in \citetalias{Barker07b}, leading us
to conclude that they are mostly caused by inaccuracies
in the stellar tracks.  However, the fit qualities
are still $\sim 2 - 4\sigma$ worse than the 
\citetalias{Barker07b} solutions.
This difference could arise from the approximations
made in producing the chemical evolution models and
the uncertainties inherent in the stellar yields.
Second, age and metallicity are no longer completely
free parameters so errors in the stellar tracks 
cannot be as easily hidden by arbitrary
combinations of age and metallicity.
Third, we do not account for a systematic change
in the metallicity spread with age.
Finally, we have not allowed for an inflow or outflow
of stars with 
other ages and metallicities into the 
single coherently evolving region we are modelling.
This mixing of stars, perhaps 
caused by gravitational interactions with spiral arms, 
could artificially increase
the spread in the age-metallicity relation
and it could mean the stars in this region originated, 
on average, at another location in M33 
\citep{Sellwood02,Roskar08}.

\begin{figure*}
\includegraphics[]{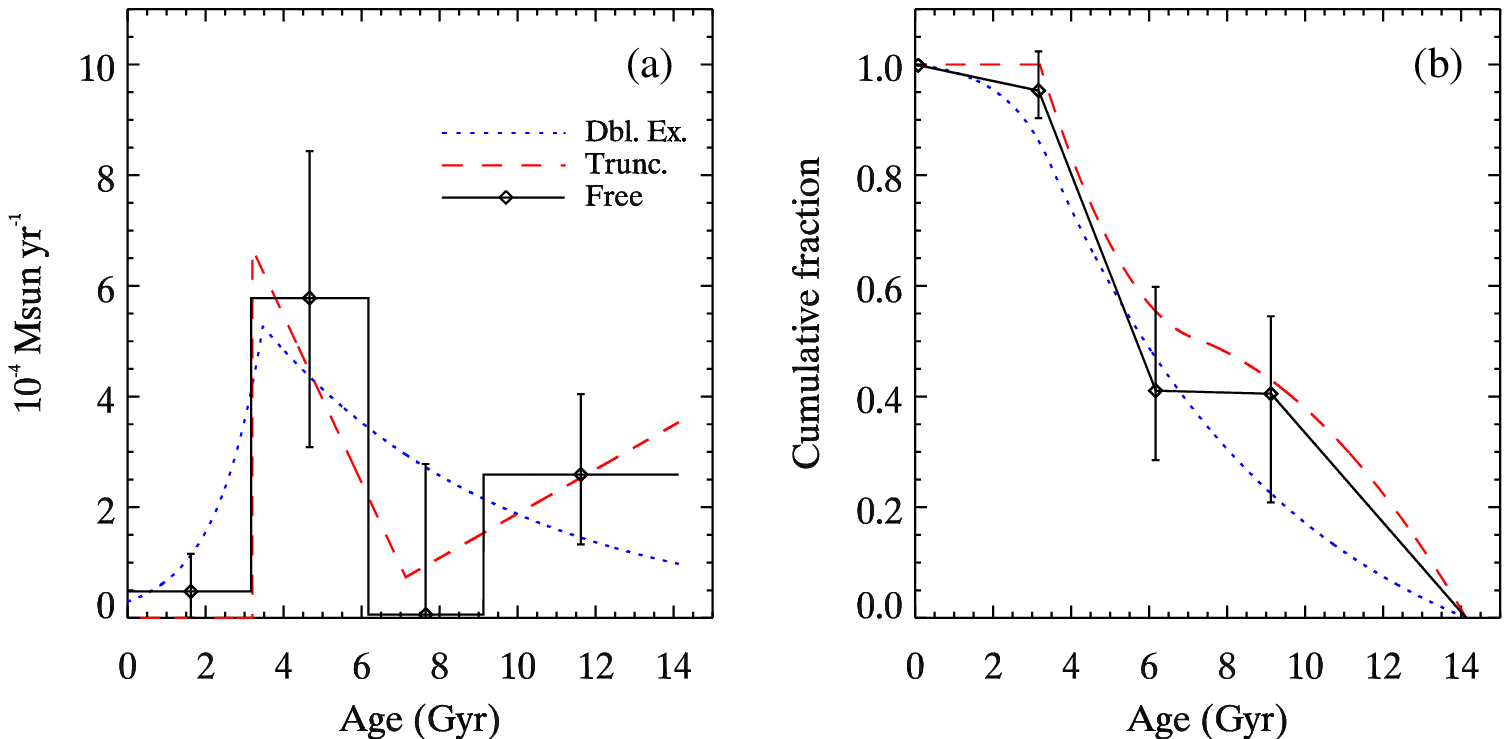}
\caption{(a) Inflow history and (b) inflow 
cumulative distribution function of double exponential (dotted line), 
truncated (dashed line), and free inflow models (solid line)
using the Padova tracks.
For clarity, only the free inflow model errors are shown.}
\label{fig:A1_padova_ira_compare_ifh}
\end{figure*}

In Fig.\ \ref{fig:A1_padova_ira_compare_sfh}, 
we summarize the distribution of stellar ages
and metallicities in the three best inflow models
for the Padova tracks.
The panels show the 
(a) SFH, (b) age cumulative distribution function (age CDF), 
(c) Z AMR, and (d) Z metallicity distribution 
function (Z MDF) of all stars ever formed.
The dotted and dashed lines are, respectively, 
the double exponential and truncated inflow models.
The solid line with diamonds and error bars
corresponds to the free inflow model.
The open circles are the results from \citetalias{Barker07b}.
Note that only the free inflow model errors are shown for clarity.
Additionally, we show the IFH and inflow CDF of these 
three models in Fig.\ \ref{fig:A1_padova_ira_compare_ifh}. 

In the best three inflow models, 
$\sim 50 - 60\%$ of the gas accretion
takes place between 3 and 7 Gyr ago.  
Anything less would produce too few intermediate-age stars.
Second, at most $\sim 10\%$ 
of the gas was accreted in the last 3 Gyr.
Any amount in excess of that would lead to a recent SFR
that is too high and produce too many  young stars
on the blue plume MS at $(V-I) \sim 0.0$.  
Third, the preferred outflow efficiency is $\approx 1 \pm 0.5$,
which is smaller than the typical values of $\sim 5 - 10$ estimated for
dwarf galaxies in the LG \citep{Lanfranchi04,Carigi06}.
Finally, the results for the SFH, age CDF, AMR, and
Z CDF are similar to the \citetalias{Barker07b} solutions.
This lends support to our conclusions in \citetalias{Barker07b}, 
and suggests they are not significantly affected by
unphysical combinations of age and metallicity.

\begin{figure*}
\includegraphics[]{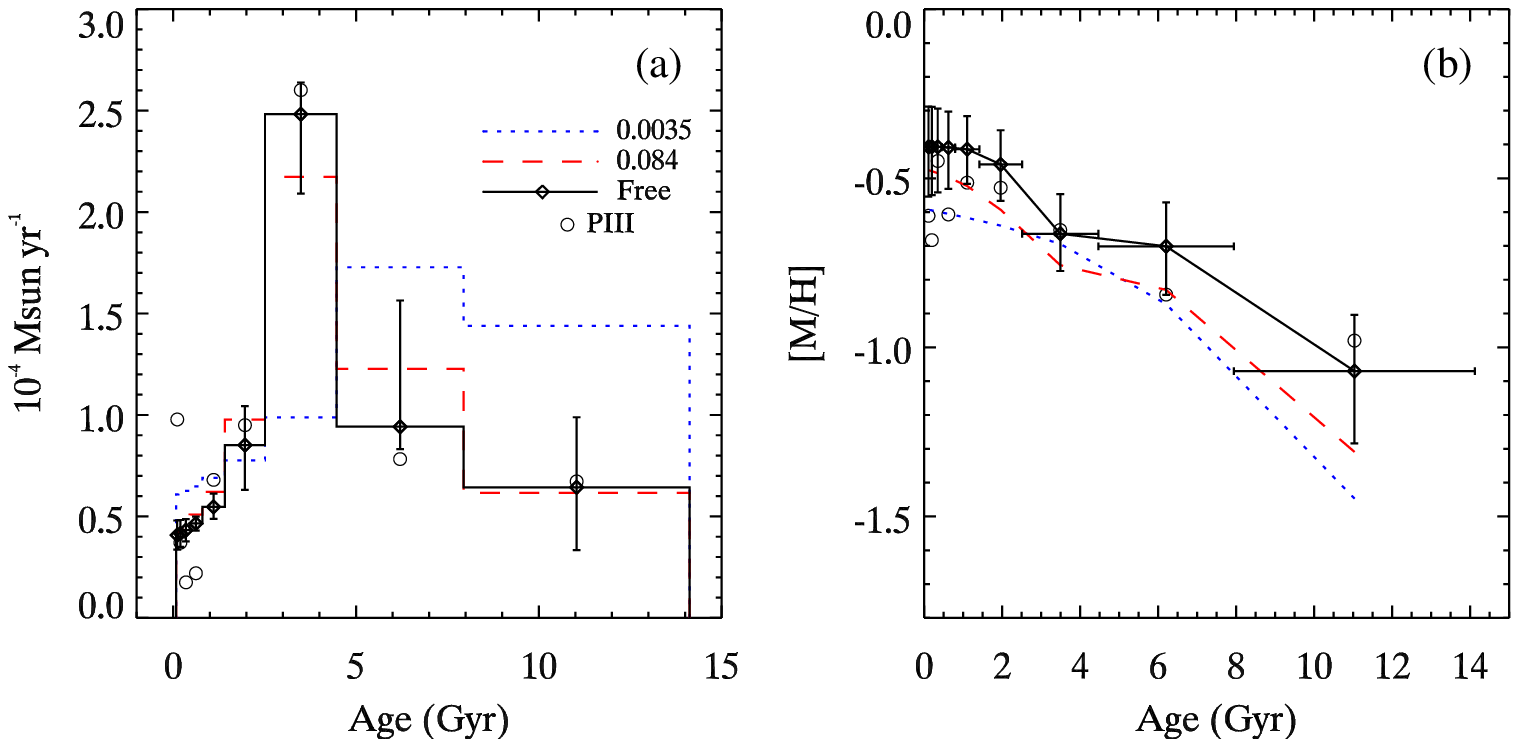}
\caption{
(a) Star formation history and (b) age-metallicity relation
using two independently estimated values for M33's 
star formation efficiency, $\epsilon = 0.0035$ (dotted line) 
and 0.084 (dashed line; see text for details).
The original free inflow model (solid line with diamonds)
treated $\epsilon$ as a free parameter, 
resulting in a best-fitting value of 0.62. 
Open circles are the results from \citetalias{Barker07b}.
Only the original free inflow model errors are shown
for the sake of clarity.
}
\label{fig:fig9_col.eps}
\end{figure*}

The preferred values for $\epsilon$ 
tend to be of order $\sim 0.1 - 1.0$.  The resulting 
gas depletion timescale, $\tau_g = \Sigma_{g}/\Sigma_{SFR}$ 
$\approx 3 - 10\ \rm Gyr$, 
is on the low end of the equivalent timescale in dwarf galaxies, 
which ranges from a few to several tens of Gyr 
\citep{Taylor05b,Karachentsev07,Calura08}.
An $\epsilon$ value as low as 0.0035 \citep{Heyer04} is strongly
disfavored because, as Fig. \ref{fig:fig9_col.eps} 
demonstrates, the resulting 
distributions of stellar age and metallicity are
skewed to higher and lower values than the \citetalias{Barker07b} solution.
If, instead, we fix $\epsilon$ to have a higher value, like 0.084, as
implied by the local current gas density and SFR in our field, then
the age and metallicity distributions are more reasonable.
This value of $\epsilon$ is only $1.2\sigma$ 
larger than the \citet{Heyer04} value and only $1.1\sigma$
lower than the best-fitting value of 0.62 in the free inflow model.
The variation in $\epsilon$ among the best three inflow models
indicates the exact parametrization of the IFH can
affect it by almost an order of magnitude.
Based on the tests in \S \ref{sec:var}, 
a similar uncertainty in $\epsilon$ arises from various assumptions
built into the chemical models, like the stellar yields
and the metallicity of the inflowing gas.

\begin{figure*}
\includegraphics[]{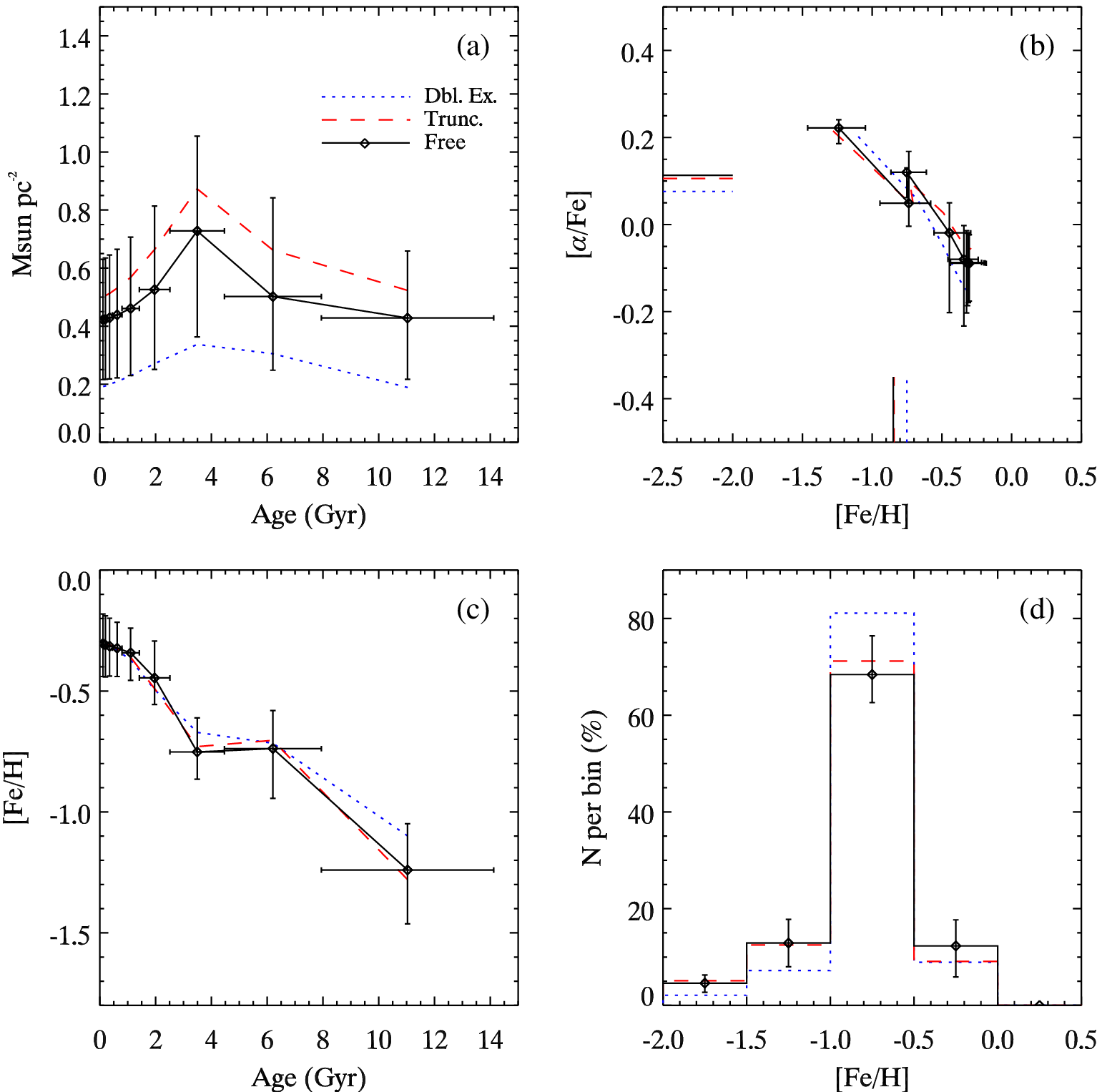}
\caption{Gas mass and chemical composition in the double 
exponential (dotted line), truncated (dashed line), 
and free inflow models (solid line) using
the Padova tracks.
(a) Evolution of the gas mass surface density, 
(b) [$\alpha$/Fe] vs.\ [Fe/H] relation, 
(c) age-metallicity relation for iron, and (d) metallicity
distribution function for iron.
Only the free inflow model errors are shown for the sake of clarity.
The horizontal and vertical line segments in panel (b)
denote, respectively, the mean [$\alpha$/Fe] and [Fe/H] of all
stars ever formed.}
\label{fig:A1_padova_ira_compare}
\end{figure*}

By incorporating the chemical evolution equations into the
CMD fitting, we can extract more
information from the solutions and make more 
predictions that can be tested against observations.
Fig.\ \ref{fig:A1_padova_ira_compare} 
summarizes some predictions for the three best inflow models.
Each figure shows (a) the gas mass 
averaged over each of the 9 SFH bins, (b) the stellar mass-weighted 
$\langle$[$\alpha$/Fe]$\rangle$ and $\langle$[Fe/H]$\rangle$ of all stars 
formed in each SFH bin, (c) the Fe AMR, 
and (d) the Fe MDF.
In panel (b), the horizontal and vertical lines
denote $\langle$[$\alpha$/Fe]$\rangle$ 
and $\langle$[Fe/H]$\rangle$ of 
all stars ever formed, respectively.  The line types are the same
as in Fig.\ \ref{fig:A1_padova_ira_compare_sfh}.

Because of the presence of gas inflow,
the gas mass rises in the early evolutionary stages
and reaches a maximum several Gyr later.  
The gas mass begins to decline as the inflow rate becomes small
but star formation and outflow continue.
Interestingly, the total mass actually decreases
in a few cases when the outflow rate exceeds the inflow rate.
The mean [$\alpha$/Fe] of the oldest age bin is $\sim 0.2 \pm 0.1$
while that of the youngest bin is $\sim -0.1 \pm 0.1$.  
Because the majority of stellar mass formed within the
oldest three age bins, $\langle$[$\alpha$/Fe]$\rangle$ of all stars
ever formed is $\sim 0.1$, even though the majority
of age bins have $\langle$[$\alpha$/Fe]$\rangle$ $< 0.0$.
Note that the [$\alpha$/Fe] vs.\ [Fe/H] relation is not always
single-valued, since [$\alpha$/Fe] can 
increase and [Fe/H] can decrease with
time depending on the precise interplay between gas flows, the SFR, 
and the SN Ia explosion rate (see \ref{sec:lmc}).

The Teramo models show an overall similar history as the Padova
models, with $\sim 50\%$ of the total inflow taking
place in the last 7 Gyr and $\la 5\%$ in the last 3 Gyr.
However, the inflow hiatus present in the Padova free inflow model
between $\sim 6$ and 9 Gyr is not present in the Teramo model.
The mean [$\alpha$/Fe] is also similar between 
the two sets of tracks.  The Teramo models predict
gas masses smaller by a factor of $\sim 2$ and 
metallicities
$\sim 0.2$ dex higher at all ages. 
This metallicity difference is somewhat 
larger here than in \citetalias{Barker07b}.
We believe this is mostly due to the
metallicity of the 6.2 Gyr age bin in the Teramo solution
now being $\sim 0.3$ dex larger than it is in \citetalias{Barker07b}.
The overall faster enrichment of the Teramo solutions 
compared to the Padova solutions 
requires higher fitted values of $\epsilon$.  
In a couple of the inflow
models, $\epsilon$ reaches its upper limit of 10.
We repeated these fits with an upper limit of 100
and the resulting fit qualities were improved 
by $\sim 1 - 2\sigma$, but the details of the SFH, IFH, 
and chemical evolution were not significantly different.

\section{Varying the Model Parameters}
\label{sec:var}

\begin{figure*}
\includegraphics[]{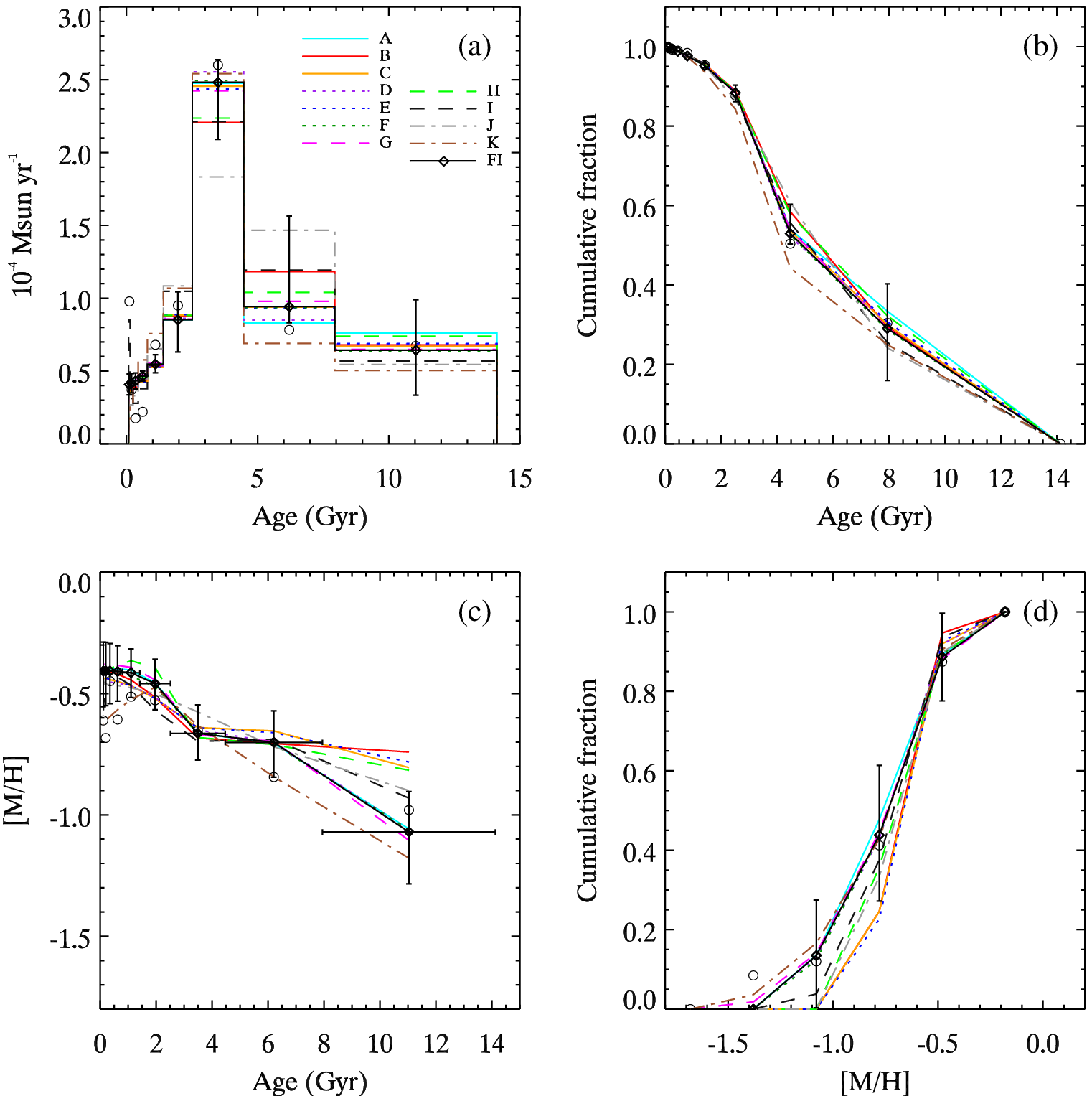}
\caption{Effects of varying the model parameters
on the star formation history and age-metallicity 
relation using the Padova tracks.  
The solid line with diamonds is the original free inflow
model and the other lines are
tests A through K (see text for details).
Open circles are the results from \citetalias{Barker07b}.
Only the original free inflow model errors are shown
for the sake of clarity.
The panels are identical to Fig.\ ~\ref{fig:A1_padova_ira_compare_sfh}.}
\label{fig:A1_padova_ira_uncert_sfh}
\end{figure*}

\begin{figure*}
\includegraphics[]{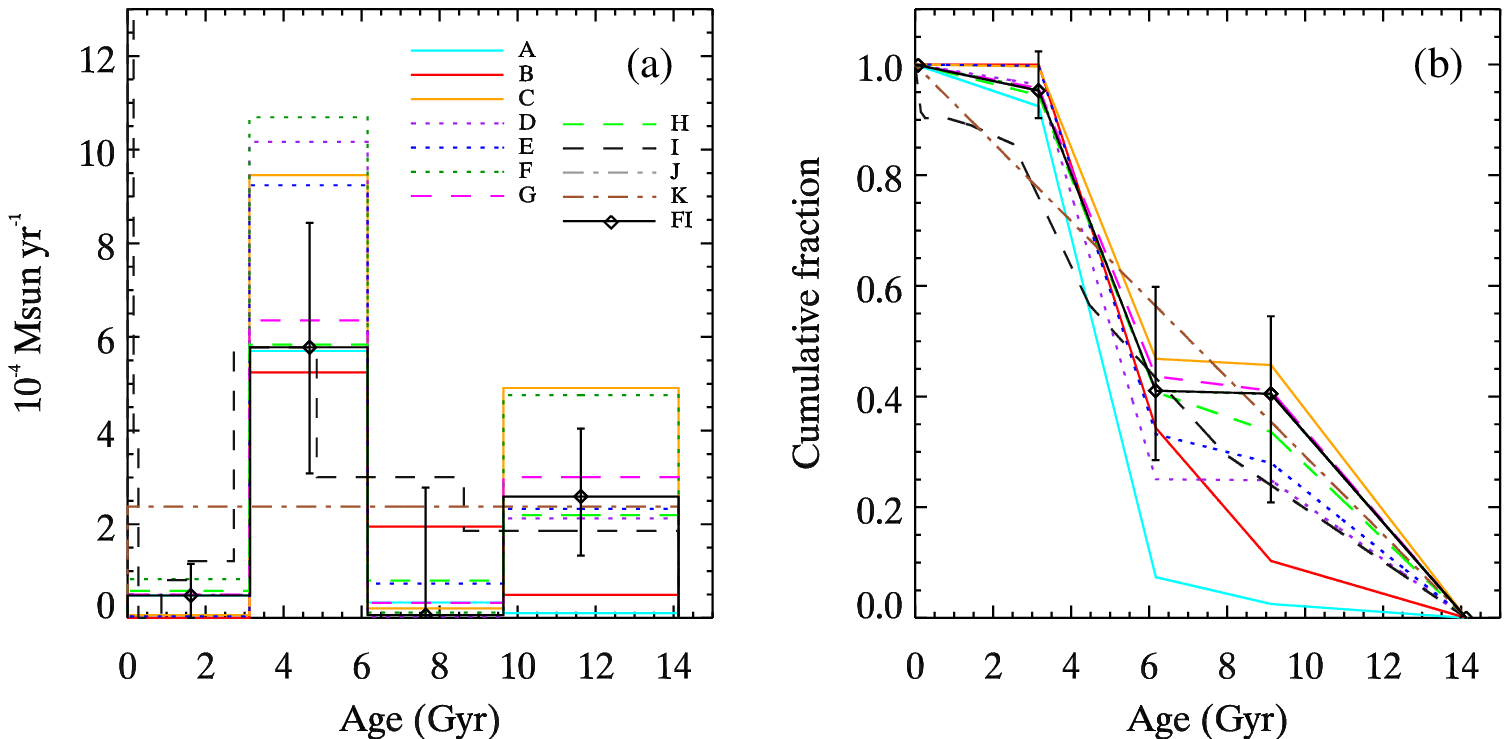}
\caption{Effects of varying the model parameters on 
the inflow history using the Padova tracks.
The solid line with diamonds is the original free inflow
model and the other lines are
tests A through K (see text for details).
Only the original free inflow model errors are shown
for the sake of clarity.
The panels are identical to Fig.\ \ref{fig:A1_padova_ira_compare_ifh}.}
\label{fig:A1_padova_ira_uncert_ifh}
\end{figure*}

\begin{figure*}
\includegraphics[]{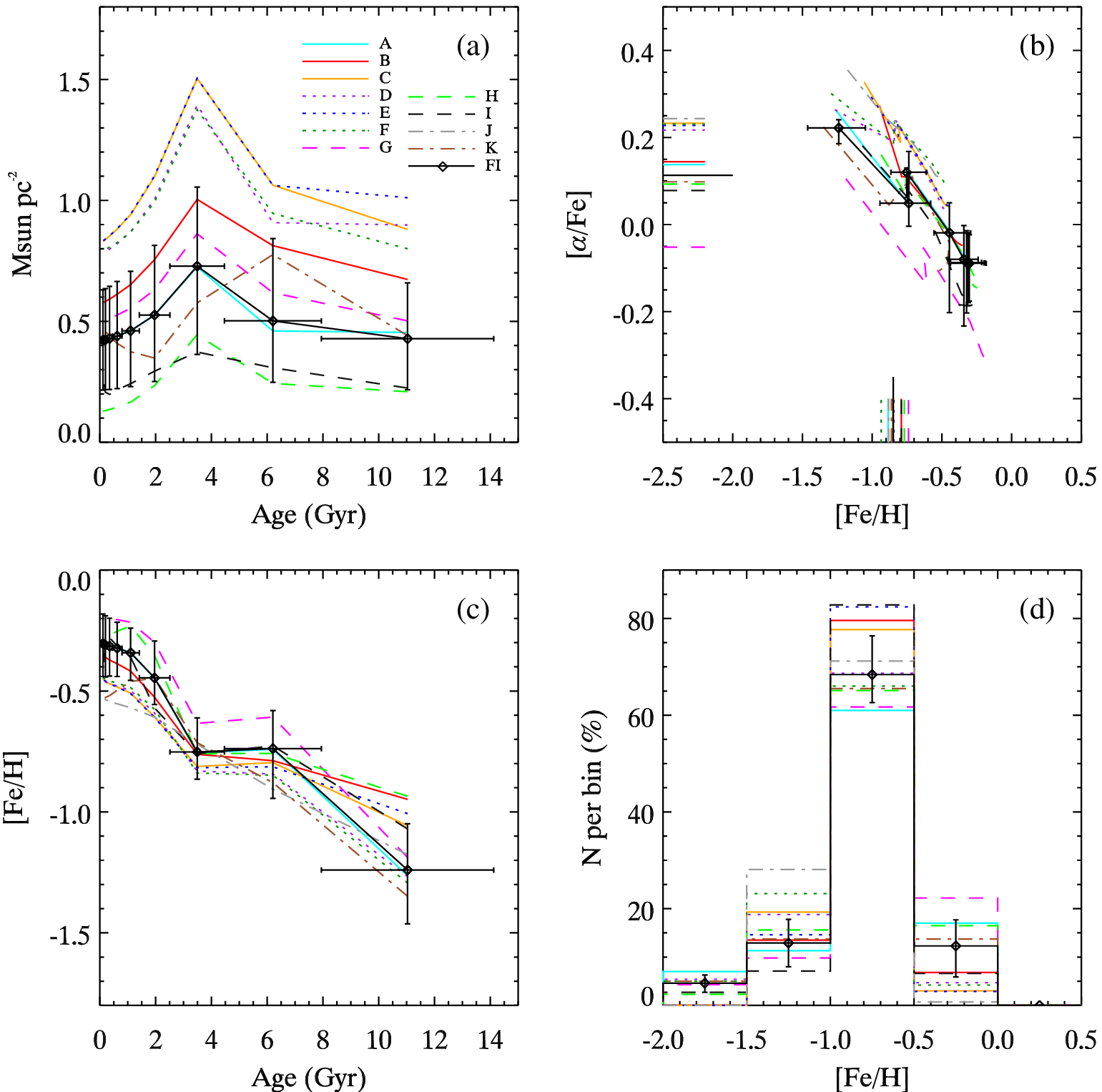}
\caption{Effects of varying the model parameters
on the gas mass and 
chemical composition using the Padova tracks.
The solid line with diamonds is the original free inflow
model and the other lines are
tests A through K (see text for details).
Only the original free inflow model errors are shown
for the sake of clarity.
The panels are identical to Fig.\ \ref{fig:A1_padova_ira_compare}.}
\label{fig:A1_padova_ira_uncert}
\end{figure*}

There were several parameters in our chemical models which we fixed
to match observations or be consistent with previous studies.
These parameters included the initial gas mass, 
initial chemical composition, composition of the inflowing gas, 
the KS relation exponent, and the stellar yields.
To estimate their effects on our results, we varied each of 
these parameters and repeated the free inflow fit.
This resulted in several new models which we refer to as A through K
and which had the following parameter changes:  
(A) an initial unenriched gas reservoir with 
$\Sigma_g(0) = 1.0\ M_{\sun}\ \rm pc^{-2}$, 
(B) the same as A but the gas reservoir 
was pre-enriched to [Fe/H] $= -1.3$, 
(C) the inflowing gas was pre-enriched by SNe~II 
to [Fe/H] $= -1.3$, 
(D) a superposition of A and C, 
(E) a superposition of B and C, 
(F) $y_i$ was increased by a factor of 2, 
(G) $y_i$ was decreased by a factor of 2, 
(H) $\kappa = 1.4$, and (I) a different binning scheme was used.
We also ran two additional tests in which $\epsilon$ was
allowed to vary with time:  
(J) a closed box model with an initial metallicity of [Fe/H] $= -1.3$ 
and (K) a free inflow model with a constant
IFR (i.e., one inflow bin rather than 4).  
While these tests are by no means exhaustive, they give a 
rough sense of the potential systematic errors that could
be introduced by the assumptions we made in our chemical models.

Figs.\ \ref{fig:A1_padova_ira_uncert_sfh} -- 
\ref{fig:A1_padova_ira_uncert} compare 
the original best-fitting free inflow model 
(line with diamonds and error bars) to
the new models (lines without error bars).
The open circles are the results
from \citetalias{Barker07b}.  Note that only the original
free inflow model errors are shown for clarity.

In general, the new results are not significantly different
from the original results and the fit qualities 
are unchanged to within $\sim 1\sigma$.  
The SFH, age CDF, MDF, and Z CDF
are the least affected by the new parameter values
and they remain close to the results of \citetalias{Barker07b}.
More variation occurs in the 
IFH, inflow CDF, gas mass evolution, and 
[$\alpha$/Fe] vs.\ [Fe/H] relation.
The IFR and gas mass in any age bin
can show variations up to a factor of $\sim 2$.
The presence of an initial gas reservoir, 
as in models A, B, D, and E, decreases the
amount of inflow required in the oldest bin because otherwise the
resulting SFR would be too high.  This means that, 
all else being equal, the 
fraction of total inflow occurring at younger ages ($\la 7$ Gyr)
in the original models is a lower limit.
Increasing [Fe/H] of the initial gas reservoir 
only increases $\langle$[Fe/H]$\rangle$ and
$\langle$[$\alpha$/Fe]$\rangle$ of the oldest age bin.  
All else being equal, increasing [Fe/H] of the inflowing gas 
shifts $\langle$[$\alpha$/Fe]$\rangle$ of all age bins 
upward $\sim 0.1$ dex and shifts the initial and final 
$\langle$[Fe/H]$\rangle$ upward and downward
by $\sim 0.2$ dex, respectively.
Models C -- F have some of the largest IFRs and gas mass densities.
In model F, a larger IFR of primordial gas
is needed to balance the increased stellar yields.
When the inflowing gas is not primordial, as in models C -- E, more of it
is required to counteract the ISM enrichment due to stellar
nucleosynthesis.


One of the largest sources of uncertainty in chemical 
evolution models is the stellar yields.
Changing the instantaneous yields, $y_i$, by a factor of 2, 
as in models F and G, shifts the [$\alpha$/Fe] vs.\ [Fe/H]
relation vertically by $\sim 0.1 - 0.2$ dex with only a small
change to the relation's tilt.
Increasing or decreasing the delayed yields, $y_{i,d}$, has the same
effect on the [$\alpha$/Fe] vs.\ [Fe/H] relation 
as decreasing or increasing, respectively, the instantaneous yields.
This fact arises because the $\alpha$-elements contribute the
majority of $y_i$ while iron contributes the majority of $y_{i,d}$.
Decreasing $y_i$ or increasing $y_{i,d}$
appear to be the only ways to significantly move
the entire [$\alpha$/Fe] vs.\ [Fe/H] relation {\it down}
relative to the original model.

To demonstrate the binning effects on the results, 
Model I has a different binning scheme from the original
free inflow solution.  Instead of 4 bins, Model I uses
the same 9 logarithmically spaced bins as the SFH.
This model is within $1\sigma$ of the original free inflow
model and has a similar inflow history and chemical evolution
to within the errors.
The two most important differences are that, first, 
the inflow hiatus appearing in the original model is no longer present.
This hiatus was possible in the original model, and not
the new one, because
of the overlap between the 
2nd oldest SFH bin and 3rd oldest IFH bin.
Second, the new binning scheme allows 
a large inflow burst in the youngest two bins ($\sim 80 - 250$
Myr) amounting to $\sim 10\%$ of the total inflow.
During this burst, the IFR increases by a factor $> 1000$
over the previous two bins.
This burst occurs in order to reproduce the apparent SFR burst
in the youngest bin seen in the \citetalias{Barker07b} solution.
As the tests and discussion in \citetalias{Barker07b} bear out, 
the reality of such a SFH burst is highly suspect 
because the SFR amplitudes in the youngest couple of age bins are 
susceptible to low number statistics.

In all the preceding results, we have attributed variations
in SFR primarily to variations in IFR because the
star formation efficiency, $\epsilon$, was constant with time.
A natural question to ask is, can the SFR variations
be due, instead, to variations in $\epsilon$?
Allowing $\epsilon$ to vary 
weakens the link between SFH and IFH,
but does not completely decouple them 
because $\epsilon$ and the IFR have 
opposite effects on the ISM metallicity.
A sudden increase in the IFR dilutes
the metals in the ISM, decreasing the enrichment rate, 
whereas a sudden increase in $\epsilon$ increases
the SFR and, hence, enrichment rate.
Phenomena like gravitational perturbations 
and stellar feedback could potentially 
change $\epsilon$, which may more generally evolve according
to local gas phase properties like 
temperature, density, pressure, chemical composition, 
molecular fraction, and interstellar radiation
field strength 
\citep[e.g.][]{Elmegreen93,SommerLarsen03,Schaye04,Schaye07,Robertson08}.

Models J and K respectively demonstrate that a closed box or constant IFR
model provide an acceptable fit to the CMD
if we allow $\epsilon$ to vary with time.
However, the amount of variation required 
(an r.m.s. deviation over time $\sim 0.5$ dex) 
may be larger than the intrinsic scatter in the
KS relation of other spirals given the vagaries of 
H$\alpha$ extinction corrections 
and the $\rm H_2$-CO conversion factor
\citep{Kennicutt98,Wong02,Komugi05,Kennicutt07,Boissier07}.
Also, the initial metallicity in model J 
must be non-zero, otherwise there are too many metal-poor
stars.
This model does a poor job of reproducing the
observed present-day gas surface density of $\sim 0.8\ M_{\sun}\ \rm pc^{-2}$
(see \S \ref{sec:obs}).  The gas density in model J 
decreases from about 3 to 2.3 $M_{\sun}\ \rm pc^{-2}$
for the Padova tracks and from about 2.2 to 1.5 $M_{\sun}\ \rm pc^{-2}$
for the Teramo tracks.
Hence, this quantity is above the range plotted
in panel (a) of Fig. \ref{fig:A1_padova_ira_uncert}.
In model K, the AMR decreases over the last 2 Gyr because
gas inflow continues while very few stars form.
Model K also has more gas inflow ($\sim 20\%$) occurring in the
last 3 Gyr than the original free inflow 
model ($< 10\%$).

\section{Comparison to Observations in M33}
\label{sec:obs}

Next, we will examine in further detail the predictions of
the original free inflow solutions, since they provide the best fits.
The Padova solutions have a present-day gas mass of 
$\sim 0.4\ \pm\ 0.2\ M_{\sun}\ {\rm pc^{-2}}$.  
From the difference between the Padova and Teramo
solutions, we estimate a systematic
error of a factor of $\sim 2$ due to uncertainties 
in the stellar tracks.
The projected mean HI column density in this field is 
$\sim 1.1\ M_{\sun}\ {\rm pc^{-2}}$, as measured from a mosaic constructed
from Very Large Array and
Green Bank Telescope observations 
(D. Thilker, private communication).  If we correct for a
disc inclination of $56\degr$ and for a helium abundance 
one-third that of hydrogen, then this becomes 
$\sim 0.8\ M_{\sun}\ {\rm pc^{-2}}$.  
The true inclination of the
HI disc at this location is somewhat uncertain because
of the well-known warp at large radii \citep{Corbelli97}. 
A change of $\pm 10\degr$ in the inclination would change the
surface density by approximately $\pm\ 0.2\ M_{\sun}\ {\rm pc^{-2}}$.
For comparison, the azimuthally-averaged HI column 
density at this radius in M33 is 
$\sim 3\ \rm M_{\sun}\ pc^{-2}$ \citep{Corbelli97,Corbelli03}.

The present-day SFR provides another 
useful check on our results.
Using GALEX near-UV and far-UV images of M33, 
Boissier et al.\ (2007; private communication) computed the UV
surface brightness in our field and then converted that
to a SFR using the relation in \citet{Kennicutt98}, which is 
a slightly modified version of the one first
presented in \citet{Madau98}.
The infrared data used to estimate extinction 
was of poor quality this far out in M33, so
the SFR could only be constrained to the range
$\sim (0.6 - 0.9) \times 10^{-4}\ M_{\sun}\ {\rm yr^{-1}}$, 
where the lower limit 
corresponds to zero extinction and the upper limit
corresponds to an extinction of $A_{FUV} = 0.49$.
The azimuthally-averaged UV SFR at this radius 
lies at the lower limit.
The chief sources of uncertainty in these limits 
are the UV flux-SFR calibration, which was based on
theoretical isochrone and spectral libraries, and
the assumption of a Salpeter IMF.  
According to \citet{Madau98},
adopting a Scalo IMF \citep{Scalo86}, which is more similar
to the IMF we have used, the resulting SFR would be
$\sim 50\%$ smaller.  Therefore, the range above
becomes $\sim (0.3 - 0.5) \times 10^{-4}\ M_{\sun}\ {\rm yr^{-1}}$, which
is in good agreement with our predictions of 
$\sim 0.4$ and $0.3 \times 10^{-4}\ M_{\sun}\ {\rm yr^{-1}}$ for the Padova
and Teramo free inflow solutions, respectively.

\begin{figure}
\includegraphics[width=3in]{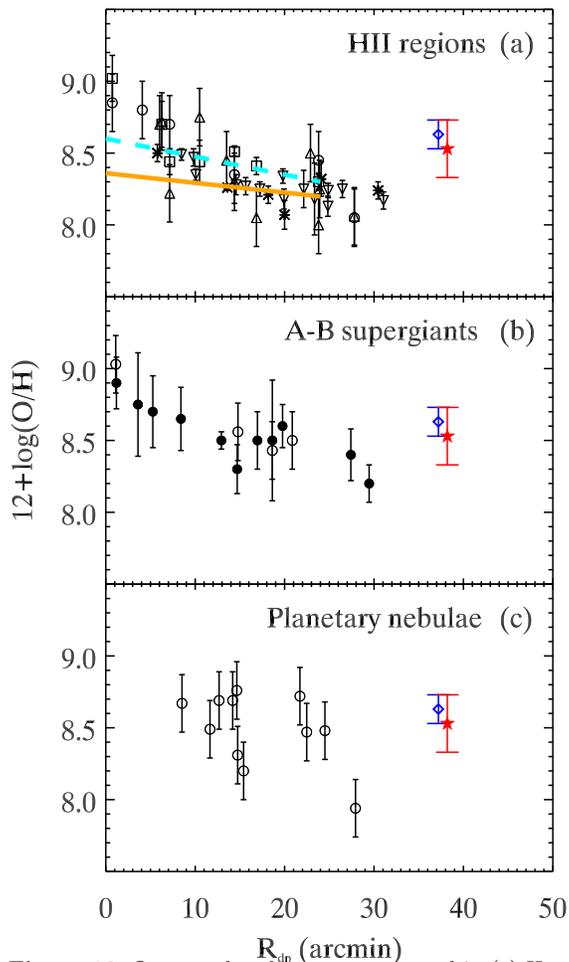}
\caption[Oxygen abundances in M33]
{Oxygen abundances as measured in
(a) \mbox{H\,{\sc ii}} regions, (b) A-B supergiant stars, and 
(c) PNe.
The solid and dashed lines in panel (a) come from 
\citet{Rosolowsky08} and \citet{Viironen07}, respectively.
In all three panels, the star and diamond are, respectively, 
the Padova and Teramo free inflow model predictions for the
present day oxygen abundance.
Their abscissa values are offset from each other for clarity.}
\label{fig:cy11_m33_HIIb}
\end{figure}

Another important check comes from oxygen abundances in M33.  
\citet{Magrini07a} compiled an extensive catalog of 
previously measured abundances 
in \mbox{H\,{\sc ii}} regions, type A-B supergiant stars, 
and planetary nebulae (PNe).
This catalog is plotted in Fig.\ \ref{fig:cy11_m33_HIIb}
with 
the predictions of the Padova and Teramo 
free inflow models for the present-day oxygen abundance 
(star and diamond, respectively), whose abscissa 
values are offset from each other for clarity.  
The \mbox{H\,{\sc ii}} regions and supergiants probe the present-day ISM 
abundance whereas the PNe probe the ISM abundance at older ages.
The masses and lifetimes of the PNe progenitors are not known with
great certainty.  The sample of Magrini et al.\ 
covers only the brightest two magnitudes of the PNe
luminosity function, so it could be biased toward the
high end of the progenitor mass range.  Therefore, 
the progenitors probably
had MS lifetimes less than a few Gyr.

Most of the \mbox{H\,{\sc ii}} region abundances in 
the Magrini et al.\ compilation
were made from direct $T_e$ measurements, generally
considered the most reliable kind.  Eight objects
have at least two independent measurements and three of those
objects have three independent measurements.
We have also
added the recent gradient derived by \citet{Viironen07} (dashed line)
for $\sim 60$ \mbox{H\,{\sc ii}} regions 
based on the $\rm log(H\alpha/$[\mbox{S\,{\sc ii}}]$\lambda\lambda6717+6731)$ vs.\ 
$\rm log(H\alpha/$[\mbox{N\,{\sc ii}}]$\lambda6583)$ diagnostic diagram.
Recently, \citet{Rosolowsky08} reported oxygen abundances
of 61 \mbox{H\,{\sc ii}} regions in the southwest region of M33
derived from the temperature sensitive emission line
[\mbox{O\,{\sc iii}}] $\lambda 4363$~\AA.  They found a shallow gradient
which we show as the solid line in Fig.\ \ref{fig:cy11_m33_HIIb}.  

Our results are more consistent 
with an overall shallow gradient or a flattening in the
gradient at large radii \citep{Magrini07a} 
than they are with a constant gradient over M33's entire disc. 
However, even if the gradient was flat beyond $R_{dp} \sim 30 \arcmin$,
our results would be several tenths of a dex higher than 
expected from the data in Fig.\ \ref{fig:cy11_m33_HIIb}.
The magnitude of this discrepancy is comparable to the
random and systematic uncertainties in our results, 
so it may not be statistically significant.
Moreover, \citet{Rosolowsky08} noted a scatter
of 0.11 dex in their sample unattributable to measurement
errors, possibly indicating chemical inhomogeneities in M33's ISM.
This fact also highlights the importance of using large samples
of objects at many position angles within the disc. 
More importantly, the data in Fig.\ \ref{fig:cy11_m33_HIIb} 
have their own systematic uncertainties.  
For example, the abundances of \mbox{H\,{\sc ii}} 
regions tend to be lower than
supergiants and PNe by $\sim 0.2$ dex.
This offset could indicate a problem in how the 
temperature and ionization structures of \mbox{H\,{\sc ii}} regions
are treated 
\citep[e.g.,][and references therein]{GarciaRojas07,Esteban04,Kennicutt03}, 
but larger samples need to be compared
before the nature of this discrepancy can be fully understood.

\section{Discussion}
\label{sec:chemdisc}

\subsection{Comparison to Models 
of the Solar Vicinity and Magellanic Clouds}
\label{sec:lmc}

\begin{figure}
\includegraphics[width=3in]{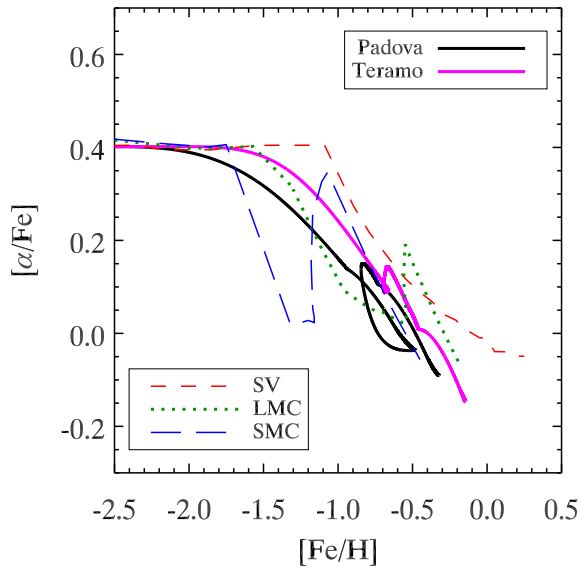}
\caption
{Comparing the [$\alpha$/Fe] vs.\ [Fe/H] relation 
for M33, the SV, LMC, and SMC.
The solid lines are the free inflow solutions while the
other lines are the chemical models in \citetalias{Pagel95}
and \citetalias{Pagel98}.}
\label{fig:cy11_afe_compare}
\end{figure}

In Fig.\ \ref{fig:cy11_afe_compare}, we compare 
the [$\alpha$/Fe] vs.\ [Fe/H] relation in this region of M33 to
that of the SV, LMC, and SMC 
derived in \citetalias{Pagel95} and \citetalias{Pagel98}.  
The models for these other three systems have
been cited often in the literature because of their simplicity 
and ability to reproduce observations fairly well.
The solid lines 
show the free inflow solutions 
using the Padova and Teramo tracks without being
averaged over each SFH bin.  
Viewing the relation in this manner facilitates comparison
with the other systems and can lead to greater physical insight, 
as long as we remember that the overall shape is most
robust.  
This last fact can be appreciated from the 
overall similarity of 
the acceptable inflow solutions 
in Fig.~\ \ref{fig:A1_padova_ira_compare}.

Fig.\ \ref{fig:cy11_afe_compare} reveals that
M33's [$\alpha$/Fe] vs.\ [Fe/H] 
relation is generally similar to the other systems.  
Because we have used almost identical chemical evolution equations
and identical stellar yields as \citetalias{Pagel95} and 
\citetalias{Pagel98}, our models 
show a similar overall behavior in response to gas flows.
The discontinuities in the relation for each 
system can be traced back to the interplay
between the SFR, IFR, and the delayed injection of iron into the
ISM from SNe Ia.  Therefore, the differences between the 
relations arise primarily from differences in the SFH and IFH.

In the IRA, the abundance ratio of any two elements is a 
constant determined by the ratio of their nucleosynthetic yields.
Thus, the initial $[\alpha/{\rm Fe}]$ ratio is 
${\rm log}(y_{\alpha}/y_{\rm Fe}) - 
{\rm log}(X_{\alpha}/X_{\rm Fe})_{\sun} \approx~ 0.4$.
After 1.3 Gyr have elapsed, the DPA begins operating as the 
first SN Ia inject iron into the ISM and 
produce the first downturn, or ``knee'', in the relation. 
The [Fe/H] location of this knee is dependent on the SFR, 
which, in turn, depends on the IFR and outflow rate.  
The SFR was on average higher in the SV than in
the other systems, so it experienced a faster enrichment
and the knee occurs at a higher metallicity.

The large loop in the Padova model results from three successive
steps:  a sudden
large increase in the IFR at an age $\sim 6$ Gyr 
which dilutes metals in the ISM
(see the IFH in Fig.\ \ref{fig:A1_padova_ira_compare_ifh}a), 
a corresponding increase in the SFR 
which raises [$\alpha$/Fe], and the subsequent drop in [$\alpha$/Fe]
1.3 Gyr later due to SN Ia.  
The Teramo model experiences a much smaller
loop because it does not have as large an increase in
the IFR at 6 Gyr.

\subsection{Implications for the Formation of the MW's Halo}

The $\Lambda$CDM hierarchical picture of galaxy formation predicts that
dark matter haloes are built up from the accretion/merging 
of smaller subhaloes similar to the 
protogalactic fragments 
proposed by \citet{Searle78}.  
This picture can be tested by comparing the 
chemical composition of the MW's stellar halo
field population to such subhaloes or their
possible present-day analogs, such as the
MW's dSph satellites.
As reviewed in detail by \citet{Geisler07} and references therein, 
the MW's field halo population is chemically distinct
from these other systems, indicating that the former 
was not built up from the latter.
In particular, 
the present-day dSphs and other LG galaxies are characterized by [$\alpha$/Fe]
ratios lower than most of the MW's halo, especially at
[Fe/H] $\ga -2$.  This is commonly 
interpreted to mean that the SFR in the
dSphs was lower and, therefore, 
their chemical enrichment was too slow 
to reach the halo's metallicity with a high 
[$\alpha$/Fe] before SNe Ia 
contributed significant amounts of iron.

\begin{figure*}
\includegraphics[]{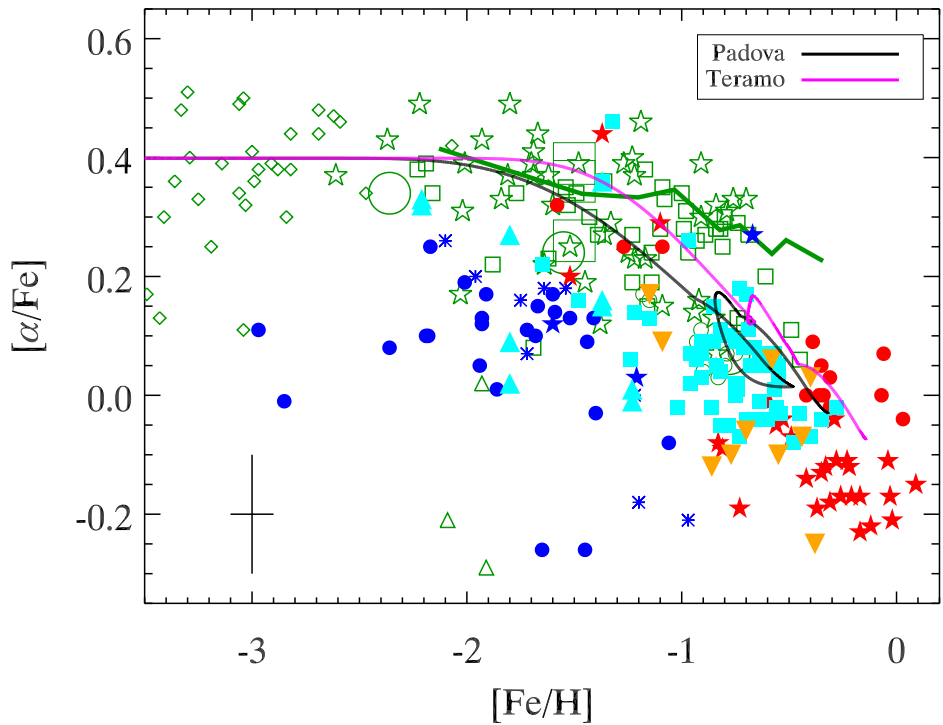}
\caption
{Comparing the [$\alpha$/Fe] vs.\ [Fe/H] relation 
in one location in M33's outskirts to spectroscopic abundance ratios 
of individual stars in various LG systems.  
The data were compiled from the literature 
and presented in \citet{Geisler07}, 
to which we refer the reader for more details.
The systems shown are the MW halo 
(green open points and green line), 
low-mass dSphs (blue filled circles, stars, and asterisks), 
Sagittarius dSph (red filled circles and stars), 
LMC (cyan filled squares and upward pointing triangles) 
and dIrrs (orange filled downward pointing triangles).
The Padova and Teramo free inflow models are 
black and magenta lines, respectively.  
The error bars in the lower left corner represent
typical random measurement uncertainties.}
\label{fig:cy11_afe_geisler}
\end{figure*}

It is interesting to compare our results for M33's outer disc
to the spectroscopic measurements in other LG systems.
Such an exercise
places our results into the context of the LG as a whole and could have some
implications for the formation of the MW's halo.
To that end, 
in Fig.\ \ref{fig:cy11_afe_geisler}, we summarize 
the high-resolution spectroscopic measurements of [$\alpha$/Fe]
and [Fe/H] in these systems.  This figure is reproduced 
from \citet{Geisler07} except that we have added
the Padova and Teramo free inflow solutions as the 
black and magenta lines~ 
\footnote{
To be consistent with \citet{Geisler07}, we plot 
[$\alpha$/Fe] = ([O/Fe]+[Mg/Fe]+[Si/Fe]+[Ca/Fe]+[Ti/Fe])/5
in Fig. \ref{fig:cy11_afe_geisler}}.
The green line represents the halo's dissipative
collapse component while the green open 
stars represent the halo's accretion component, 
as defined by \citet{Gratton03} based on orbital kinematics.
These authors postulated the dissipative component
formed during the collapse 
of the MW's gaseous halo, 
as earlier outlined by \citet{Eggen62}, and the accretion
component originated in the accreted subhaloes.

The green open diamonds at [Fe/H] $\sim -3$ 
represent extremely metal-poor halo 
giants analysed by \citet{Cayrel04}, for which no separation
into kinematic components has been made.
All other green open points 
represent samples of MW halo
stars selected for their unusual orbital properties
and which therefore are additional candidates for originating
in chaotically accreted subhaloes.
The other systems represented in Fig.\ \ref{fig:cy11_afe_geisler} 
are low-mass dSph satellites of the MW 
(blue filled circles, stars, and asterisks), 
Sagittarius dSph 
(red filled circles and stars), 
LMC (cyan filled squares and upward pointing triangles), 
and several dIrrs (orange filled downward pointing triangles).
Typical random measurement uncertainties are indicated
in the bottom left corner.  
We refer the reader to \citet{Geisler07} 
for a more detailed explanation
and for a list of all the
references for the data.

Based on our results, if we could obtain high resolution spectroscopic
abundances for the location we have studied in M33, we would expect
most of its stars to lie near $\langle$[$\alpha$/Fe]$\rangle \sim 0.1$ 
and $\langle$[Fe/H]$\rangle \sim -0.7$.
Of all the systems presented in Fig.\ \ref{fig:cy11_afe_geisler}, 
this region of parameter space is most similar to the LMC.
We also note that M33 resembles the MW's halo 
more than the low-mass dSphs.
The plateau at [Fe/H] $\la -2.0$ was built into the
models via the chemical yields (see \S\ref{sec:lmc}), but the key fact is 
that M33 maintains a higher [$\alpha$/Fe] ratio at a given [Fe/H]
relative to the low-mass dSphs.
The agreement between M33 and the MW's halo becomes progressively
worse at [Fe/H] $\ga -2$.
However, M33's relation overlaps with some of the 
candidate accretion stars suggesting they 
could have originated 
in a system whose SFH, AMR, and IFH were 
similar to M33's outer disc, though not necessarily a 
spiral galaxy like M33.  This is consistent with the suggestion of 
\citet{Geisler07} that some of the accreted MW halo stars
originated in high mass dwarf systems.

Although we have little information
on M33's {\it inner} disc evolution, the presence of a metallicity
gradient \citep{Magrini07a} and a wide variety
of ages inferred from CMDs \citep{Sarajedini00}
hint at an extended SFH with a faster
enrichment taking place there.  Therefore, we
hypothesize that the knee in the [$\alpha$/Fe] vs.\ [Fe/H] 
relation in M33's {\it inner} disc occurs at a higher metallicity
than in the outer disc.  This brings us to another key point, that
candidate accretion halo stars with different
[$\alpha$/Fe] but the same [Fe/H] did not necessarily
originate in different objects.  
In other words, the $\alpha$-element and iron abundances alone
do not necessarily provide enough information to tag 
individual halo stars to unique progenitors \citep{Freeman02}.
This is because the inner regions of a 
protogalactic fragment could have become enriched with metals faster
than the outer regions.  Therefore, the candidate accretion 
stars that we observe today
with low and high [$\alpha$/Fe] could have originated in the 
outer and inner regions of the fragment, respectively.

How realistic is such a scenario?
How much time would have been necessary 
and available for this hypothetical
fragment to attain the range of [Fe/H] and [$\alpha$/Fe] 
observed today in the MW halo's accreted stars?
Our results indicate M33's outskirts took 
$\sim 4$ Gyr to reach [Fe/H] $\ga -1.0$ and [$\alpha$/Fe] $\sim 0.1$.
\citet{Zentner03} used a semi-analytic model to 
examine the survival of orbiting subhaloes in the MW's
potential.  
They found that for a typical subhalo orbit, the survival
time after entering the MW's halo ranged 
from 3 to over 8 Gyr, depending on the 
subhalo's mass and concentration relative to the MW's.
Any time spent outside the MW halo prior to accretion 
would only relax the 
necessary post-accretion survival time.
Thus, it seems entirely
plausible that this hypothetical fragment could have had enough 
time to develop a population gradient before disruption.

Present-day population gradients have been found in most of the dSphs 
\citep[e.g.,][]{Harbeck01,Winnick03,Tolstoy04,Koch06,
Stetson98,Bellazzini05,Battaglia06,Faria07,Komiyama07,
McConnachie07}, 
in the Sagittarius stream \citep{Bellazzini06,Chou07}, 
NGC 6822 \citep{deBlok06}, and in M33 
\citep[][\citetalias{Barker07a}, \citetalias{Barker07b}]{Rowe05}.
In most cases, the inner regions of the galaxies are
more metal rich and/or older than the outer regions.
In another recent study, \citet{Bernard08} found that
RR Lyrae stars in the central region of Tucana were 
more metal-rich than in the outer region, suggesting a
metallicity gradient was already present in the first few Gyr of that
galaxy's history.

The evolution of metallicity gradients in disc galaxies
is a matter of some debate, with some studies predicting
the gradient to increase with time and others predicting 
the opposite \citep[][and references therein]{Magrini07a}.  
The evolutionary model of M33 presented
in \citet{Magrini07a} predicts the metallicity gradient
to have been steeper in the past.  
Measurements of the spatially resolved SFH and AMR of these
low mass galaxies could help us understand if they 
can develop chemical gradients in the first few Gyr of their lives
or if such gradients were present in the gas from which they formed.

\section{Conclusions}
\label{sec:chemconc}

In the framework of a simple chemical evolution
scenario which adopts instantaneous and delayed recycling
for the nucleosynthetic products of Type II and Ia SNe, we have modelled 
the observed CMD in one location in M33's outskirts.  
In this scenario, interstellar gas forms stars at a rate
modulated by the KS relation 
and gas outflow occurs at a rate proportional to the SFR.
Compared to the common CMD fitting method of allowing age and
metallicity to be free parameters, this scenario yields 
a more physically self-consistent SFH with fewer
free parameters and makes more predictions that
can be tested with independent observations.  
Moreover, this method puts broad constraints on the role of gas 
flows and extends the method of chemical 
fingerprinting to stellar systems, like M33, that are beyond the reach
of current high resolution spectrographs.

The precise details
of our results depend on which stellar 
evolutionary tracks are used, Padova or Teramo.  
Nevertheless, the broad trends appear to be robust.
We found that, when the star formation efficiency, 
$\epsilon$, is constant, the canonical closed box model fails to
reproduce the observed distribution of stars in the CMD.
Models with an exponentially declining IFR from 14.1 Gyr ago 
exhibit similar discrepancies, but to 
a lesser magnitude.  Instead, the inflow models which
best reproduce the observed CMD have
a significant fraction ($\ga 50\%$) 
of gas inflow taking place in the last 7 Gyr 
and a smaller fraction ($< 10\%$) taking place
within the last 3 Gyr.  
This leads to a SFH in overall agreement with what
we found in \citetalias{Barker07b}, and suggests the
traditional method of synthetic CMD fitting can be 
physically self-consistent.

Allowing $\epsilon$ to vary
with time, as might be expected if the star
formation time-scale depends on local ISM properties, 
significantly improves 
the closed box model if the initial metallicity
is [Fe/H] $\sim -1.3$, but the resulting 
present-day gas mass surface density is
too high compared to the observed value.
A model with a varying $\epsilon$ and constant, non-zero IFR
reproduces the CMD and gas mass about as well as the
best constant $\epsilon$ models.  
In this case, more inflow can take place in the last 3 Gyr, 
but there is still a significant fraction ($\sim 50\%$) 
of inflow occurring over the last 7 Gyr.
The amount of variation in $\epsilon$ required
by these models, however, could be larger than the intrinsic
scatter in the KS relation of other galaxies.

We also examined the [$\alpha$/Fe] vs.\ [Fe/H] relation, 
a key diagnostic of the evolutionary history of stellar systems.
Like the MW's dwarf satellites, the bulk of stars at this
location in M33's outskirts have [$\alpha$/Fe] ratios lower than
the MW's halo field.
Stars formed over 8 Gyr ago 
have $\langle$[$\alpha$/Fe]$\rangle$ $\sim 0.2 \pm 0.1$
and stars forming today have 
$\langle$[$\alpha$/Fe]$\rangle$ $\sim -0.1 \pm 0.1$.
The mean [$\alpha$/Fe] ratio of all stars ever formed 
at this location in M33 was
found to be $\sim 0.1 \pm 0.1$.
From the tests we have conducted, 
we estimate that the systematic errors of these values
are $\sim \pm 0.2$ dex, comparable to that of
some high resolution spectroscopic measurements of other systems 
\citep[e.g.][]{Monaco05,Geisler07}.



Our results paint a picture in which M33's outer disc
formed from the protracted inflow of gas over several Gyr 
with at least half of the total inflow
occurring since $z \sim 1$, but relatively little since $z \sim 0.25$.
All the acceptable inflow models have a similar 
IFR $\sim 2 \times 10^{-4}\ \rm 
M_{\sun}\ yr^{-1}\ kpc^{-2}$ 
averaged over the lifetime of the Universe.
This estimate is uncertain by at least a factor of 2
and the IFR could have intrinsic variations of a factor of $\sim 3$
when averaged over 2 $-$ 3 Gyr time-scales.
Nevertheless, it still gives a rough indication of
the mean gas IFR that could occur in
the outskirts of low mass spiral galaxies.  

A useful baseline for comparison is the lifetime 
average IFR in the SV, predicted by
\citet{Colavitti08} to be 
$\sim 4 \times 10^{-3}\ \rm\ M_{\sun}\ yr^{-1}\ kpc^{-2}$.
This estimate comes from several chemical evolution
models which adopt different shapes for the IFH 
and which reproduce many local observables, like the
abundances of various elements, 
MDF of long-lived stars, and present-day 
gas fraction, total mass density, SFR, IFR, and SN rate.
That the mean IFR in the SV has been higher than in M33's
outer disc is not surprising since the MW is more
massive than M33 and the SV is at a much 
smaller radius (in terms of disc scale lengths) 
than the M33 field we have studied.

\citet[][]{SommerLarsen03} ran an N-body
cosmological simulation and selected 12 dark matter
haloes for more detailed simulations
which included baryons.
Two Milky Way-type spirals, S1 and S2, that resulted from the simulations 
experienced gas accretion at a mean rate of $\sim 10^{-3}\ M_{\sun} \rm\ 
yr^{-1}\ kpc^{-2}$ at radii of $6 - 7$ disc scale lengths.
This rate is about 5 times larger than the mean
IFR we have found at a similar location in 
M33's outer disc, but S1 and S2 
were about 5 times more massive than M33.
The total disc averaged accretion rates of 
the least massive simulated galaxies, with rotation velocities
comparable to M33's, were 
$\sim 15 - 35\%$ smaller than those of S1 and S2
at $z = 0$.
Therefore, if the {\it outer} disc accretion rate 
scales with a galaxy's rotation velocity in the same way as
the {\it total} disc averaged rate, 
and if this scaling holds for all redshifts, 
then our results are
consistent with the simulations of \citet[][]{SommerLarsen03}.

What is the source and nature of the gas inflow?
If it occurs in the disc plane, it could be caused by spiral density waves, 
viscosity in the disc gas, or gas with lower
angular momentum falling onto the disc at 
larger radii and flowing toward M33's nucleus
\citep{Lacey85b,Lin87,Bertin96b,Portinari00,Roskar08}.
Gas flowing from above the disc plane could come from 
the condensation of a hot halo corona as described in \S \ref{sec:intro}.
However, such a process is expected to be more efficient
in a massive spiral like the MW than in a late-type spiral
like M33 \citep{Dekel06}.

The condensed clouds predicted by 
numerical simulations to fall onto a disc galaxy 
have properties similar to the MW's HVC population \citep{Peek08}.
Recent surveys of HI emission around M31 and M33
have revealed an analogous HVC population around these galaxies
\citep{Thilker04,Braun04,Grossi08}.  
\citet{Thilker04} detected 20
clouds within 50 kpc of M31's disc and 
the maps presented by \citet{Grossi08}  
show many within $\sim 20$ kpc 
of M33 with an inferred total mass 
$\geq 5 \times 10^7\ M_{\sun}$.
The M31/33 clouds appear to have a mean mass 
$\sim 10^5\ M_{\sun}$ and size $\sim 1$ kpc 
\citep{Westmeier05,Grossi08}.
The lifetime integrated inflow mass of $\sim 10^6\ M_{\sun}$ 
in the M33 region we have studied requires the accretion
of $\sim 10$ such clouds.  
Extrapolation to M33's entire disc is risky given
how little we know about its evolution and the population
of clouds around M33.
In addition to condensed halo gas, the clouds
could also be 
low-mass dark matter subhaloes that never formed stars or 
the tidal debris from recent mergers and 
interactions \citep{Thilker04,Grossi08}.
In the future, we plan to apply the techniques presented in
this paper to other regions of M33 and other galaxies.
This approach can provide insights on the 
origin and amount of accreted gas and coupling 
(or lack thereof) between the baryonic and dark matter 
accretion histories of these systems.

\section*{Acknowledgments}

We would like to thank the anonymous referee for
constructive feedback which helped improve the
content and clarity of this paper.
We thank Fred Hamann and Stephen Gottesman for stimulating discussions
and Annette Ferguson for useful comments on a draft.
We gratefully acknowledge Doug Geisler, Samuel Boissier, and David Thilker for
sharing their data with us.
This paper is based on observations made with the NASA/ESA Hubble Space
Telescope, obtained at the Space Telescope
Science Institute, which is operated by the Association of
Universities for Research in Astronomy, Inc., under NASA contract
NAS 5-26555.  These observations are associated with program GO-9479.
Support for program GO-9479 was provided by NASA through a
grant from the Space Telescope Science Institute.
This work has made use of the IAC-STAR Synthetic CMD computation code. 
IAC-STAR is supported and maintained by the computer division of the 
Instituto de Astrof\'{i}sica de Canarias.
We acknowledge the University of Florida High-Performance 
Computing Center for providing computational resources and support 
that have contributed to the research results reported within this paper.


\bibliographystyle{mn2e}

\bibliography{references}

\bsp

\label{lastpage}


\end{document}